\documentclass[a4paper,11pt]{article}
\pdfoutput=1
\usepackage{jheppub}

\usepackage[utf8]{inputenc}

\usepackage{amssymb,amsmath,amsfonts}
\usepackage{mathtools}
\usepackage{mathrsfs}
\usepackage{bbm}
\usepackage{slashed}
\usepackage{nicefrac}
\usepackage{graphicx}
\usepackage{caption}
\usepackage{subcaption}
\usepackage{esint}
\usepackage{graphicx}
\usepackage[dvipsnames]{xcolor}
\usepackage{array}
\usepackage{simplewick}
\usepackage{hyperref}
\usepackage{xparse}
\usepackage{xspace}
\usepackage{tikz}
\usetikzlibrary{decorations.pathmorphing}
\usetikzlibrary{automata,positioning}
\usepackage{cancel}
\usepackage[normalem]{ulem}
\usepackage{xifthen}
\usepackage{dsfont}
\usepackage[titletoc]{appendix}
\usepackage{booktabs}
\usepackage{units}

\def\del{\partial}

\newcommand{\eqn}[1]{Eq.~\eqref{#1}}

\long\def\comment#1{ }

\newcommand{\nn}{\nonumber\\ }

\def\be{\begin{eqnarray*}}
\def\ee{\end{eqnarray*}}
\def\beq{\begin{eqnarray}}
\def\eeq{\end{eqnarray}}

\def\0{{\boldsymbol 0}}
\def\k{{\boldsymbol k}}

\def\q{{\boldsymbol q}}

\def\bell{{\boldsymbol \ell}}

\def\btheta{{\boldsymbol \theta}}

\def\rme{{\rm e}}
\def\rmd{{\rm d}}

\def\and{ \quad\text{and}\quad}

\def\cC{{\cal C}}

\def\cP{{\cal P}}
\def\cK{{\cal K}}

\def\cO{{\cal O}}

\begin{document}

\title{Non-linear dynamics of jet quenching }

\author[a]{Yacine Mehtar-Tani,}
\emailAdd{mehtartani@bnl.gov}
\affiliation[a]{Physics Department, Brookhaven National Laboratory, Upton, NY 11973, USA}

\date{\today}

\abstract{We develop a comprehensive analytic framework for jet quenching in QCD media, based on a medium-induced parton cascade sourced by collinear virtual splittings. We show that the energy flow out of the jet cone, driven by turbulent gluon cascades, is governed by a non-linear rate equation that resums gluon splittings at arbitrary angles and is enhanced by the medium length, $L$. The solution of this equation sets the initial condition for a non-linear DGLAP-like evolution equation, which describes the collinear early vacuum cascade resolved by the medium at angles exceeding the medium resolution angle, $\theta_c$. For asymptotic jet energies, the medium-induced cascade displays an exponential behavior that generalizes the Poisson-like distribution of parton energy loss. This formulation enables the resummation of leading contributions in $\alpha_s \ln (1/R)$, and $\alpha_s \ln (R / \theta_c)$, and powers of $\alpha_s L$. We briefly explore the limit of strong quenching, where analytic treatments are feasible, offering insights into the impact of parton cascades on jet quenching. These results provide guidance for future numerical simulations and analytical investigations. }


\date{\today}
\maketitle
\flushbottom

\section{Introduction}\label{sec:intro}

The observation of quenching of high-$p_T$ hadrons \cite{Adler:2002tq, Adler:2002xw,Adcox:2001jp} and fully reconstructed jets at RHIC and LHC \cite{Aad:2012vca,Khachatryan:2016jfl,Aad:2014bxa,Adam:2015ewa} has prompted extensive efforts to develop a comprehensive theoretical framework for jet quenching in a dense QCD medium. These efforts aim to account for multiple scattering and parton shower effects \cite{Arnold:2008zu, Mehtar-Tani:2012mfa, Blaizot:2012fh, Blaizot:2013vha, Blaizot:2015lma, Mueller:2016gko, Sievert:2018imd, Caucal:2018dla, Caucal:2019uvr, Arnold:2020uzm, Caucal:2020zcz, Mehtar-Tani:2021fud, Takacs:2021bpv, Mehtar-Tani:2019tvy, Mehtar-Tani:2019ygg, Barata:2020sav, Barata:2021wuf, Andres:2020vxs, Vaidya:2020cyi, Schlichting:2020lef, Mehtar-Tani:2018zba, Adhya:2019qse}, as well as coherence effects arising from quantum interference or its suppression \cite{Mehtar-Tani:2010ebp, Mehtar-Tani:2011hma, Mehtar-Tani:2011vlz, Casalderrey-Solana:2011ule, Mehtar-Tani:2011lic, Mehtar-Tani:2012mfa, Casalderrey-Solana:2012evi, Mehtar-Tani:2017ypq, Mehtar-Tani:2017web, Caucal:2019uvr, Caucal:2018dla,Abreu:2024wka}. More recently, advancements have included addressing the contributions of flowing and inhomogeneous media \cite{Andres:2022ndd, Sadofyev:2021ohn, Barata:2022krd, Barata:2023qds}. This framework seeks to quantitatively describe the experimental data and uncover the novel QCD dynamics that govern the propagation and fragmentation of highly virtual partons within a hot and dense QCD medium. Beyond the iconic jet nuclear modification factor \cite{Mehtar-Tani:2021fud}, various jet substructure studies \cite{Larkoski:2014wba,CMS:2017qlm,Chien:2016led,Mehtar-Tani:2016aco,Caucal:2021bae,Caucal:2019uvr,Barata:2023bhh}—conducted both analytically and using Monte Carlo event generators—have revealed intriguing qualitative features of medium-modified jets \cite{ALICE:2021aqk,ALICE:2022hyz,Song:2023sxb}. To advance the field further, it is essential to develop precision theoretical tools that enable systematic analytic computations of jet observables in heavy-ion collisions. 

The purpose of this work is to develop a comprehensive framework that unifies the various physical processes previously studied in the literature, which have often been addressed in a fragmented manner with limited insight into the overarching all-order structure of jet observables. A significant milestone was recently achieved with the derivation of a factorization formula for the inclusive jet cross-section \cite{Mehtar-Tani:2024smp}. The present work builds upon and extends this result.

The primary mechanism driving jet quenching is parton energy loss, which occurs through elastic scattering and radiation processes. The latter, gluon radiation, becomes equally important in large media, as the coupling constant factor associated with radiation is enhanced (on average) by the square of the medium length, $L$ \cite{Wang:1991xy,Gyulassy:1990ye,Wang:1992qdg,Gyulassy:2000fs,Baier:1994bd,Baier:1996kr,Zakharov:1996fv,Zakharov:1997uu,Gyulassy:2000er,Wiedemann:1999fq,Baier:2001yt,Wiedemann:2000za,Arnold:2002ja,Salgado:2003gb}. Due to the non-local nature of the radiative process, medium-induced gluon emission can be triggered coherently by multiple scatterings during the emission (or formation) time. This coherence results in the Landau-Pomeranchuk-Migdal (LPM) suppression of the radiation spectrum at high frequencies, corresponding to long formation times. The corresponding radiation spectrum is given by:
\beq\label{eq:rad-spect}
\omega \frac{\rmd I}{\rmd \omega} \equiv \frac{\alpha_s C_R}{\pi} \sqrt{\frac{\omega_c}{\omega}} \,,
\eeq
for formation time scales $ t_f = \sqrt{ \omega/\hat{q}} \ll L $ (which corresponds to $\omega \ll \omega_c \equiv \hat{q} L^2 $). For frequencies higher than \( \omega_c \), the spectrum is more strongly suppressed, scaling as $1/\omega $ \cite{Mehtar-Tani:2019tvy}. Here, $\hat q$ is the jet quenching parameter which is a transport coefficient that characterizes the diffusion in the medium of energetic partons in transverse momentum space, i.e., $\langle k_\perp^2\rangle \sim \hat q L$.

In a regime where the medium length and scattering center density are sufficiently large, the probability of multiple emissions increases significantly \cite{Baier:2001yt,Jeon:2003gi}. This leads to the formation of medium-induced cascades \cite{Blaizot:2013vha,Blaizot:2013hx,Blaizot:2014ula,Blaizot:2014rla}, which are understood to be the dominant mechanism for transporting energy in a self-similar way from the high-energy jet down to the medium's soft scale, such as the temperature. The kinetic equation that describe the medium induced cascade effectively resums powers of $(\alpha_s L/t_f)^n$, which arise from the iterative application of the radiative spectrum \eqn{eq:rad-spect}, in the regime where $\ell_{\rm mfp} \ll t_f \ll L$ \cite{Blaizot:2013vha}.

The inclusive properties of the turbulent gluon cascade have been extensively studied over the past decade \cite{Blaizot:2014ula,Blaizot:2014rla,Caucal:2019uvr,Mehtar-Tani:2018zba,Schlichting:2020lef,Mehtar-Tani:2022zwf,Soudi:2024yfy}, with substantial progress made at NLO \cite{Arnold:2020uzm,Arnold:2021pin,Arnold:2022fku,Arnold:2022mby,Arnold:2023qwi,Arnold:2024whj,Arnold:2024bph} including the resulting renormalization of the jet quenching parameter \cite{Blaizot:2014bha,Caucal:2022mpp,Ghiglieri:2022gyv,Caucal:2022fhc,Arnold:2021pin,Arnold:2021mow,Caucal:2021lgf,Blaizot:2019muz,Iancu:2018trm,Mehtar-Tani:2017ypq,Wu:2014nca,Iancu:2014sha,Liou:2013qya,Iancu:2014kga}. However, due to the jet definition—where partons are reconstructed within a specific solid angle—a complete picture of jet energy loss requires understanding how energy flows out of the jet cone via parton cascades. This is not a straightforward problem, as gluons radiated outside the cone can subsequently radiate back inside, as illustrated in Fig.~\ref{fig:cascade}, leading to a complex interplay between radiation inside and outside the jet region. This behavior is reminiscent of the processes responsible for generating large logarithms in non-global observables \cite{Banfi:2002hw}.

In this article, we develop an all-orders analytic description of the energy flow away from a jet region centered around the parent parton. Our main result is a non-linear evolution equation that resums medium length powers $\alpha_s L$, which we then extend to include the collinear, virtuality-driven vacuum cascade, which resums powers of $\ln (1/R)$ corresponding to out of cone DGLAP evolution, as well as, powers of $\ln (R/\theta_c)$ which relate to intra-jet color charges  that are resolved by the medium down to a resolution angle $\theta_c\sim (\hat q L^3 )^{-1/2}\ll 1$.  As a result of destructive interference, splittings below this critical angle are not resolved and therefore don't affect jet energy loss \cite{Mehtar-Tani:2011hma,Mehtar-Tani:2012mfa,Mehtar-Tani:2017ypq,Casalderrey-Solana:2011ule,Mehtar-Tani:2017ypq}.   The vacuum cascade develops prior to medium-induced processes which set the initial condition to the non-linear DGLAP cascade. Furthermore, we recover and generalize the non-linear evolution equations previously derived for jet quenching weights in the limit of soft gluon radiation.

This article is organized as follows: In Section~\ref{sec:eloss-dist}, we establish the general factorization formula for the inclusive jet cross-section and introduce the normalized energy loss distribution. In Section~\ref{sec:med-shower}, we derive the non-linear master equation governing the evolution of the energy loss distribution through the medium-induced gluon cascade. In Section~\ref{sec:vacuum}, we discuss the impact of the early collinear cascade on energy loss, described by a non-linear DGLAP equation. Section~\ref{sec:steep-spect} provides an analytical analysis of the jet spectrum, highlighting that in the strong quenching regime, multi-subjet configurations are significantly suppressed by a Sudakov-like factor. Finally, in Section~\ref{sec:NP-effects}, we address the inclusion of soft physics occurring at the medium temperature scale through a non-perturbative factor, enabling a simple analytic treatment of medium response. Finally, we address the generalization to higher orders in Section~\ref{sec:higher-orders} and present our conclusions in Section~\ref{sec:concl}.

\section{Jet energy loss distribution}\label{sec:eloss-dist}

Consider an energetic gluon with energy (longitudinal momentum) $E\equiv p^+ $ propagating through the quark-gluon plasma (QGP)\footnote{We adopt the following convention for the light-cone variables: $p^+=(p^0+p^3)/2$. }. This gluon may either be the jet initiator or a gluon emitted from an already developed parton cascade centered around the jet axis taken to be the z-axis.

At the initial time $ t_0 $, we assume the gluon’s momentum forms a small angle $ \theta \ll 1 $ with the z -axis. In the small-angle approximation, the dynamics are confined to the transverse plane. Thus, we define the two-dimensional vector $ \btheta \sim (\theta, \phi) $, where $ \phi $ is the azimuthal angle,
\beq
\btheta= \frac{\k}{E}\,.
\eeq
Here, $\k$ is the initial gluon transverse momentum.

The jet is defined by all particles that fall within the domain $\cC_{\rm in}$ in solid angle, typically characterized by an opening angle $R$. The jet region is thus defined as $\cC_{\rm in} = \{ \btheta \in \mathbb{R}^2 \, | \, |\btheta| < R \}$.

We want to compute the energy loss distribution $S_{\rm loss}(\epsilon)$  of the initial gluon longitudinal momentum $E$ that loses an amount $\epsilon$ out of the jet cone,
\beq\label{eq:excl-amp}
S_{\rm loss}(\epsilon, t, t_0, E,\btheta) = \sum_{m=1}^\infty   \prod_{i=1}^m  \int\frac{\rmd k^+_i \rmd^2 \k_i}{2(2\pi)^3 k_i^+ } |A_m(k_1,...,k_m,p;t,t_0)|^2 \delta(\epsilon-  \bar n \cdot k_{\rm out} ) \Theta_{\rm alg}  \,,\nn
\eeq
where $\Theta_{\rm alg}$ denotes phase space constraints due to the jet clustering algorithm and $\bar n \cdot k_{\rm out}= k^+_{\rm out}=\sum\limits_{i=1}^{m} k^+_{i,\rm out}$ represents the total longitudinal momentum of the collinear partons located in the `out' region at the time $t$, with $\bar n\equiv(1,0,0,-1)/2$, a null vector that defines the light cone direction opposite to the jet direction of propagation, and $A_m$ is the exclusive amplitude for an initial parton $p$ at $t_0$ fragmenting into $m$ collinear partons at time $t$. The delta function represents the measurement of energy lost by particles with momentum that are measured outside the jet  region $\cC_{\rm in}$.  In \eqn{eq:excl-amp}, momentum conservation is implicit.

By construction, the energy loss distribution is normalized to 1 as a consequence of unitarity,
\beq \label{eq:norm}
\int_0^{+\infty } \rmd \epsilon \, S_{\rm loss}(\epsilon) =1\,.
\eeq
\subsection{Factorization at Leading Order}
To compute the inclusive cross-section we need the function $S_{\rm loss}$ evaluated at $t_0=0$, $t=+\infty$ and $\btheta=0$, i.e.,   $S_{\rm loss}(\epsilon, E) \equiv  S_{\rm loss}(\epsilon,+\infty, 0, E,\btheta=0)$, which we then convolve with the LO inclusive jet cross-section as an illustration (we shall generalize to all orders shortly):
\beq\label{eq:factorization-0}
\frac{\rmd \sigma_{\rm incl}}{\rmd p_T}  = \int_0^{+\infty}\rmd E\,\int_0^{+\infty} \rmd \, \epsilon\,\,  \delta(E-\epsilon- p_T)\, S_{\rm loss}(\epsilon, E)\,  \frac{\rmd \sigma^{\rm LO}_{\rm incl}}{\rmd E}  \,.
\eeq
Here, we implicitly assume that the parent leading parton defines the jet direction, though this is not strictly accurate, as jet definition algorithms depend on the final state. However, in the high-energy limit where the jet energy far exceeds the lost energy, the effects of the jet algorithms are power suppressed. Under this assumption, the jet algorithm effectively acts on the initial collinear parton cascade. A detailed discussion of relaxing this approximation is deferred to future work.

Introducing the quenching weight as a Laplace Transform of the energy loss distribution,
\beq  \label{eq:eloss-LT}
Q_\nu(E) \equiv \int^{+\infty}_0 \rmd \epsilon \,  S_{\rm loss}(\epsilon, E)\,\rme^{-\nu \epsilon  } \,,
\eeq
with the inverse transform,
\beq
S_{\rm loss}(\epsilon, E) = \int_{c-i\infty}^{c+i\infty}\frac{\rmd \nu}{2\pi i}\, Q_\nu(E)\, \rme^{\nu \epsilon}\,,
\eeq
where the $\nu$ integral runs vertically to the real axis. The value $c$  (the real part of $\nu$) is chosen such that it is greater than the real parts of all singularities of the integrand. Note that, we may equally use the Mellin representation by transforming w.r.t. to the rescaled variable $y=\epsilon/E$. The reason the Laplace transform is more convenient stems from the fact that the medium-induced part of energy loss distribution is independent of the jet energy for large values of $E$. Moreover, in the non-linear regime discussed in the next section, the evolution equation is local in $\nu$.

\eqn{eq:factorization-0} can be rewritten as
\beq
\frac{\rmd \sigma_{\rm incl}}{\rmd p_T}  = \int_0^{+\infty}\rmd E\, \int \frac{\rmd \nu}{2\pi i} \, \rme^{ (E-p_T) \nu}Q_\nu(E) \,  \frac{\rmd \sigma^{\rm LO}_{\rm incl}}{\rmd E} \,.
\eeq
We observe that in the case where the jet loses all of its energy to the plasma we have the limiting result:
\beq
\frac{\rmd \sigma_{\rm incl}}{\rmd p_T}  \to \delta(p_T)  \int_0^{+\infty}\rmd E \frac{\rmd \sigma^{\rm LO}_{\rm incl}}{\rmd E} \,.
\eeq
In order to connect our formulation with the so-called jet function that appears in the factorization of the jet cross-section in vacuum \cite{Kang:2016mcy,Dasgupta:2014yra}, we make the change of variables
\beq
p_T = x E\,.
\eeq
We then obtain
\beq\label{eq:factorization-LO}
\frac{\rmd \sigma_{\rm incl}}{\rmd p_T}  =p_T \int_0^{1}\frac{ \rmd x}{x^2}\, \int \frac{\rmd \nu}{2\pi i} \, \rme^{ \frac{(1-x)}{x}p_T \nu}Q_{\nu}(p_T/x) \,  \frac{\rmd \sigma^{\rm LO}_{\rm incl}(E=p_T/x)}{\rmd E} \,.
\eeq

\subsection{The jet function and energy loss}
In terms of the jet function \cite{Kang:2016mcy}, which can be interpreted as an energy loss distribution even in vacuum, we have
\beq\label{eq:jet-function}
J\left(x,E\right) =E\int \frac{\rmd \nu}{2\pi i} \, \rme^{  (1-x)E \nu}\, Q_{\nu}\left(E\right)  \, .
\eeq
Then, introducing the factorization scale $\mu$, the general form of \eqn{eq:factorization-LO} reads
\beq\label{eq:factorization}
\frac{\rmd \sigma_{\rm incl}}{\rmd p_T}  = \int_0^{1}\frac{\rmd x}{x}\,J(x,E,R,\mu)  \,  H(E=p_T/x,\mu)\,.
\eeq
where $H(p_T,\mu)\sim  \rmd \sigma_{\rm incl}(\mu)/\rmd p_T $, the hard matrix element that is related to jet cross-section evaluation to a factorization scale $\mu$. This formula factorizes collinear jet dynamics from the hard scattering.

Assuming the presence of single medium scale, $E_{\rm med}$, the jet function can be written as a function of dimensionless variables:
\beq
J(x,E,R,E_{\rm med},\mu)\,\rightarrow   \,J\left (x, \frac{\mu}{E R},\frac{E_{\rm med}}{E},R \right)  \, \underset{ \rm vacuum}{\rightarrow}  \,  J\left (x, \frac{\mu}{E R}\right)\,,
\eeq
where the extra $R$ dependence is due to power corrections related to medium dynamics and can be dropped in a first approximation.

For completeness, let us recall here the operator definition of the jet function for quark initiated jet in vacuum \cite{Kang:2016mcy}:
\beq
J(x, E,\mu) \equiv  \frac{x}{2 N_c}   {\rm Tr}\left[ \frac{\slashed{n}}{2} \langle 0|  \delta\left(x-\frac{ \bar n\cdot p_J}{E}\right) \chi(0)  \, |\, XJ  \rangle   \langle  \, XJ |  \bar\chi(0)|  0\rangle \right]\,,\nn
\eeq
where $p^+_J =\bar  n \cdot p_J$, or more precisely $p^+_J = p_T \cosh \eta\sim p_T$ and  $\chi$ is the collinear fermion field (similar formula can be written for collinear gluon fields) introduced in the context of Soft-Collinear Effective Field Theory (SCET)\cite{Bauer:2002nz,Bauer:2001ct,Bauer:2000yr,Bauer:2003mga,Bauer:2002aj}. The state $|JX \rangle$ represents the final-state unobserved particles $X$ and the observed jet $J$. In the medium, we simple replace the vacuum state by a medium state $|0 \rangle \to | {\rm med}\rangle$.

In this unified framework, the jet function can be interpreted as a jet energy loss distribution even in vacuum. However, a key distinction arises between the vacuum and medium cases. In vacuum, only collinear splittings at angles larger than $R \ll 1$ produce significant logarithmic contributions, i.e., $\alpha_s \ln (1/R) \sim 1$. Since only the total transverse momentum ($p_T$) of the jet is measured, the observable is insensitive to fluctuations in the jet substructure \cite{Kang:2016mcy,Dasgupta:2014yra}. As a result, splittings within the jet cone do not affect the inclusive observable and must cancel out due to unitarity.

In contrast, a QCD medium resolves different substructure fluctuations, leading to differential energy loss that impacts the observable. For example, a jet that splits into two subjets will experience more energy loss than one that does not split, causing an imbalance between real and virtual contributions. This effect manifests as a Sudakov-type suppression, first noted in \cite{Mehtar-Tani:2017web}, resumming terms of the form $(\alpha_s \ln (R/\theta_c))^m$ (with $m=0,1,...,+\infty$) where $\theta_c$ represents the medium's angular resolution scale. When $\theta_c > R$, the medium is unable to resolve the jet's finer color structure, and the result converges to the vacuum case, apart from an overall quenching factor corresponding to the total color charge. As we shall show, this mismatch will result in a non-linear evolution when collinearly enhanced contributions are resummed to all orders in perturbation theory.

When $\theta_c$ is small but not much smaller than $R$, it may become necessary to resum soft logarithms of the form $(\alpha_s \ln (p_T/E_{\rm med}))^m \sim (\alpha_s \ln (n))^m$, where $n \gg 1$ denotes the power-law index of the jet spectrum. In such scenarios, a recently developed effective field theory (EFT) based factorization approach offers a systematic framework for separating hard collinear modes associated with virtuality-induced splittings. These splittings lose energy in the medium through soft gluon emissions, which are encoded in multi-Wilson-line correlators \cite{Mehtar-Tani:2024smp}.

\begin{figure}
\centering
\includegraphics[width=9cm]{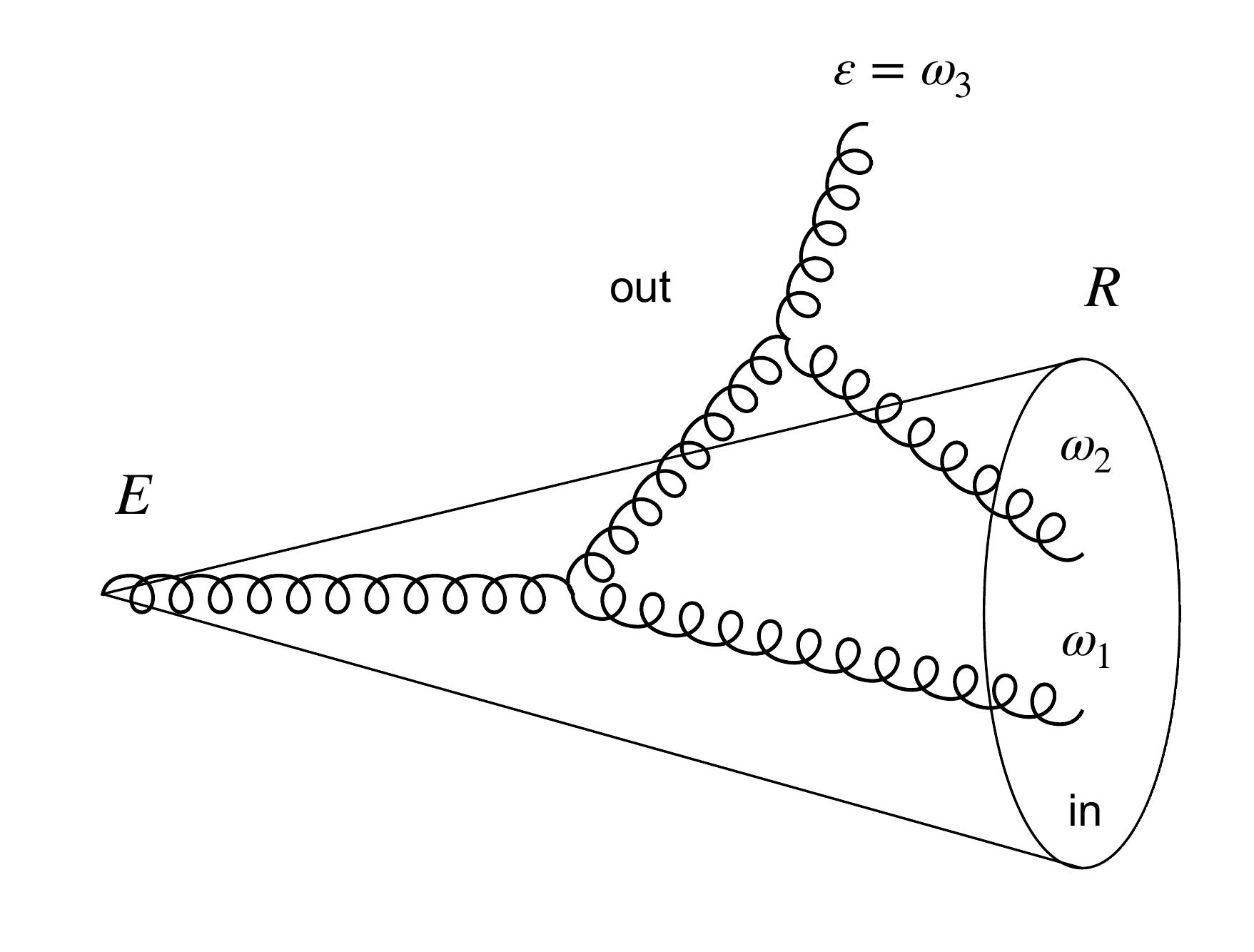}
\caption{An illustration of a high-energy gluon with energy $E$ splitting into two gluons: one that remains within the jet region $\mathcal{C}_{\rm in}$ with an opening angle $R$, and another that propagates outside the jet region before radiating back into it. This type of non-linear dynamics resembles QCD processes that generate large non-global logarithms in vacuum \cite{Dasgupta:2001sh,Banfi:2002hw}.} \label{fig:cascade}
\end{figure}

\section{Non-linear evolution driven by medium-induced cascades}\label{sec:med-shower}
\subsection{The elastic contribution}
Using the parton shower language, we shall first discuss the effect of medium induced-cascade on jet energy loss, as it is the main aspect that sets the medium and vacuum cases apart. We will then discuss the case of the virtuality ordered  vacuum cascade in Section.~\ref{sec:med-shower}.
The question we want to address here is, given a gluon with energy $E$ located initially at an angle $\btheta$ at the time $t_0$, what is the energy distribution that flows out of a roughly circular region $\cC_{\rm in}$ of radius $R$ and centered around the origin $\btheta=0$ (see Fig.~\ref{fig:cascade}).

\begin{figure}
\centering
\includegraphics[width=10cm]{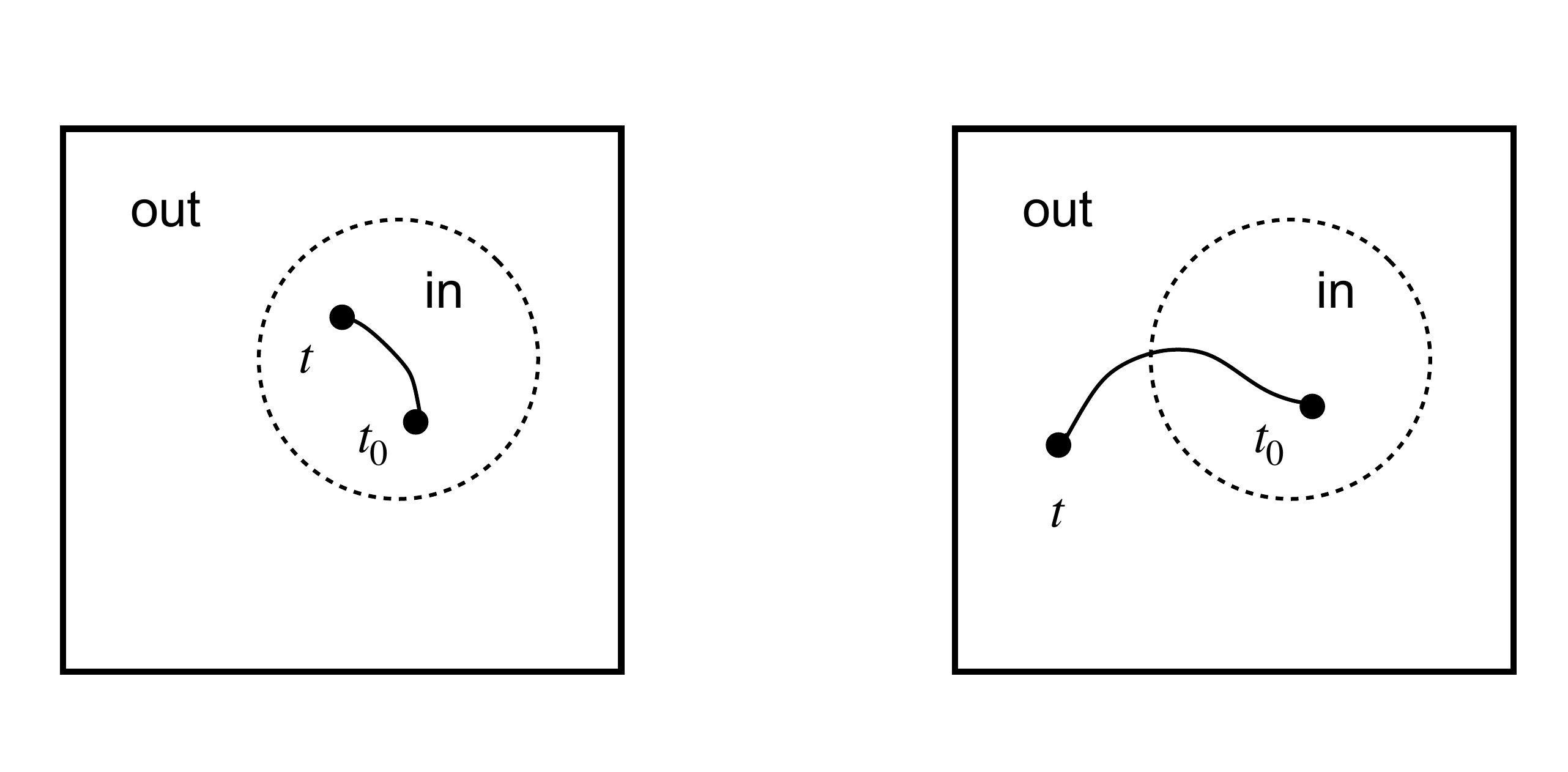}\caption{The dynamics of collinear parton evolution can be described on the transverse plane $\btheta=\q/E$. The two figures depict the two terms in the elastic ($p_\perp$-broadening) contribution to energy loss \eqn{eq:S-elastic}: the left figure corresponds to no energy loss at time $t$, since the gluon ends up inside the jet cone for which $|\btheta|<R$, while, the right figure corresponds to the case $|\btheta|<R$ where the gluon is registered outside the jet region $\cC_{\rm in}$.  }\label{fig:TMB}
\end{figure}

In the simpler case of a gluon that undergoes only elastic scattering the energy loss distribution reads
\beq\label{eq:S-elastic}
S_{\rm el}(\epsilon, t, t_0;E, \btheta)  = \int \frac{\rmd^2 \q}{(2\pi)^2}\, \cP(\q-E \btheta,t,t_0) \left[ \Theta(|\q|<RE)\, \delta(\epsilon)+\Theta(|\q|>RE)  \,\delta(E-\epsilon) \right]\,,\nn
\eeq
where $\q$ and  $\k=E\btheta$ are the final and initial gluon transverse momenta, respectively. A depiction of the two contributions is given in Fig.~\ref{fig:TMB}. The first term corresponds to the scenario where the gluon ends up inside the jet cone, resulting in no energy deposition outside the cone, i.e., $\epsilon = 0$. The second term accounts for the opposite case, where the gluon exits the jet region, leading to a total out of cone energy of $\epsilon = E$. 

The distribution $\cP(\q-\k,t,t_0)$ describes the probability for a gluon to acquire a transverse momentum kick of $\q - \k$ due to interactions with the medium between an initial time $t_0$ and final time $t$. In the independent multiple-scattering approximation it obeys the kinetic equation \cite{Blaizot:2013vha}:
\begin{align}\label{eq:el-evolution}
\frac{\del}{\del t}\cP(\k,t,t_0) = \int_\bell \cC(\bell,t)  \cP(\k-\bell,t,t_0)
\end{align}
with
\beq\label{eq:el-kernel}
\cC(\bell,t)  = 4\pi \alpha_s N_c n(t) \left[ \gamma(\bell) - (2\pi)^2\delta^{(2)}(\bell) \int_{\bell'}\gamma(\bell') \right]\,,
\eeq
where $\gamma(\bell) \simeq g^2 / \bell^4$ denotes the leading-order elastic cross-section and $n(t)$ is the density of scattering centers. Throughout, we use the shorthand notation $\int_\bell \equiv \int \rmd^2 \bell/(2\pi)^2$.

In order to make contact with the jet quenching parameter, we expand \eqn{eq:el-evolution} in gradients, for $\bell \ll \k$. The leading contribution yields a diffusion equation
\begin{align}\label{eq:el-evolution-diff}
\frac{\del}{\del t}\cP(\k,t,t_0) = \frac{1}{4}\frac{\del^2}{\del \k^2}  \hat q(\k,t) \cP(\k,t,t_0)
\end{align}
where the jet quenching parameter reads
\beq
\hat q(\k,t) =  \int_\bell \bell^2 \cC(\bell,t) \simeq  4\pi N_c \alpha_s^2 n(t) \ln \frac{\k^2}{m_D^2}\,.
\eeq

Evidently, in vacuum we have $\cP(\k-\bell,t,t_0)  \to (2\pi)^2\delta^{(2)}(\k-\bell)$, implying  $S_{\rm el}(\epsilon, t, t_0;E, \btheta=0)\to \delta(\epsilon)$.
\begin{figure}
\centering
\includegraphics[width=6cm]{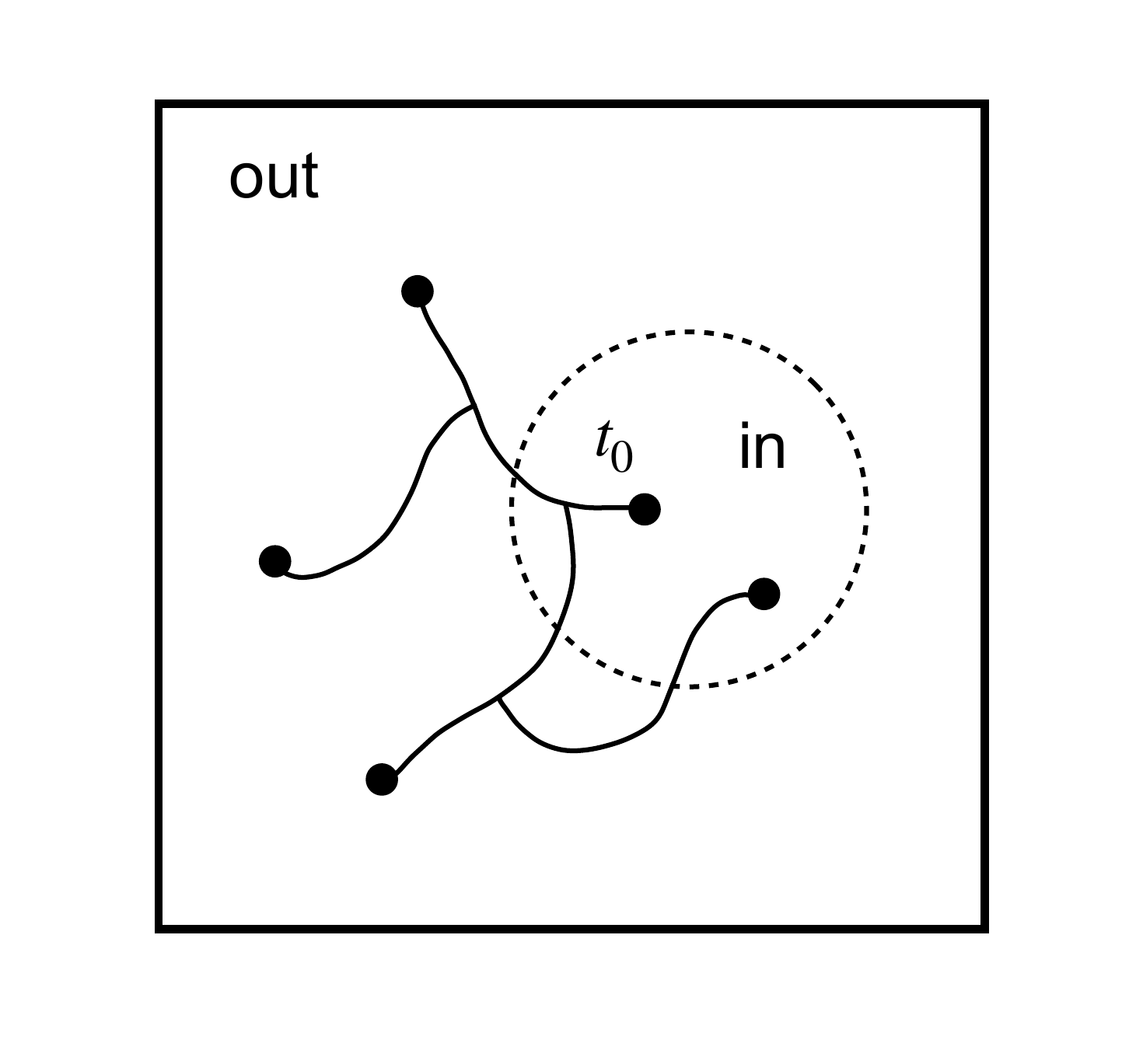}\caption{Illustration of the effect of medium-induced splitting on the evolution of energy flow out of the cone. Note that a gluon can flow out of the cone at intermediate time but then radiates a gluon back inside the cone leading to fluctuating energy loss as a consequence of the evolution of the parton cascade. In contrast to the vacuum cascade successive gluon branchings are not ordered in angles.  }\label{fig:cascade-2d}
\end{figure}

\subsection{The master equation }

Interactions with the QGP can cause the gluon to branch, initiating a gluon cascade that redistributes the initial energy among many quanta (see Fig.~\ref{fig:cascade-2d}). Therefore, in addition to elastic processes, we must account for multiple gluon branchings. During such a process, multiple scatterings cause a gluon with energy $E$ to split into two gluons with energies $zE$ and $(1-z)E$, respectively, at a rate given by $\cK(z,E)$, which, in the multiple soft-scattering approximation based on the harmonic approximation takes a simple analytic form \cite{Blaizot:2013vha},
\begin{align}\label{eq:inel-rate}
\cK(z,E)= \frac{p_{gg}(z)}{2\pi} \sqrt{\frac{(1-z+z^2) \, \hat q }{z(1-z)E}}\,,
\end{align}
where
\begin{align}
p_{gg} (z)= N_c \left[\frac{1-z}{z}+\frac{z}{1-z}+z(1-z)\right]\,,
\end{align}
is the the unregularized Altarelli-Parisi gluon to gluon splitting function \cite{Altarelli:1977zs}.

This inelastic rate describes the quasi-instantaneous collinear splitting of a gluon. Consequently, no transverse momentum is acquire during this process which is valid in the approximation where $t_f/t \to 0$. The effect of radiation on transverse momentum broadening was shown to be enhanced by a double logarithm that can absorbed into a renormalization of the elastic collision rate \cite{Blaizot:2014bha,Iancu:2014kga,Wu:2014nca}.

Details on the computation of the rate \eqn{eq:inel-rate}, along with generalizations that incorporate the hard Coulomb tail and finite-size effects, can be found in the following references \cite{Mehtar-Tani:2019tvy,Mehtar-Tani:2019ygg,Barata:2020sav,Barata:2020rdn,Barata:2021wuf,Caron-Huot:2010qjx,Andres:2023jao,Andres:2020vxs}. Additional information on the non-perturbative component of the collision rate is available in \cite{Moore:2021jwe,Schlichting:2021idr}, and for an expanding medium in \cite{Adhya:2019qse,Caucal:2020uic,Soudi:2024yfy}. This medium-induced parton shower can be viewed as a limiting case of a kinetic theory description of the plasma dynamics \cite{Arnold:2002zm,Schlichting:2020lef,Mehtar-Tani:2022zwf}.

\begin{figure}
\centering
\includegraphics[width=10cm]{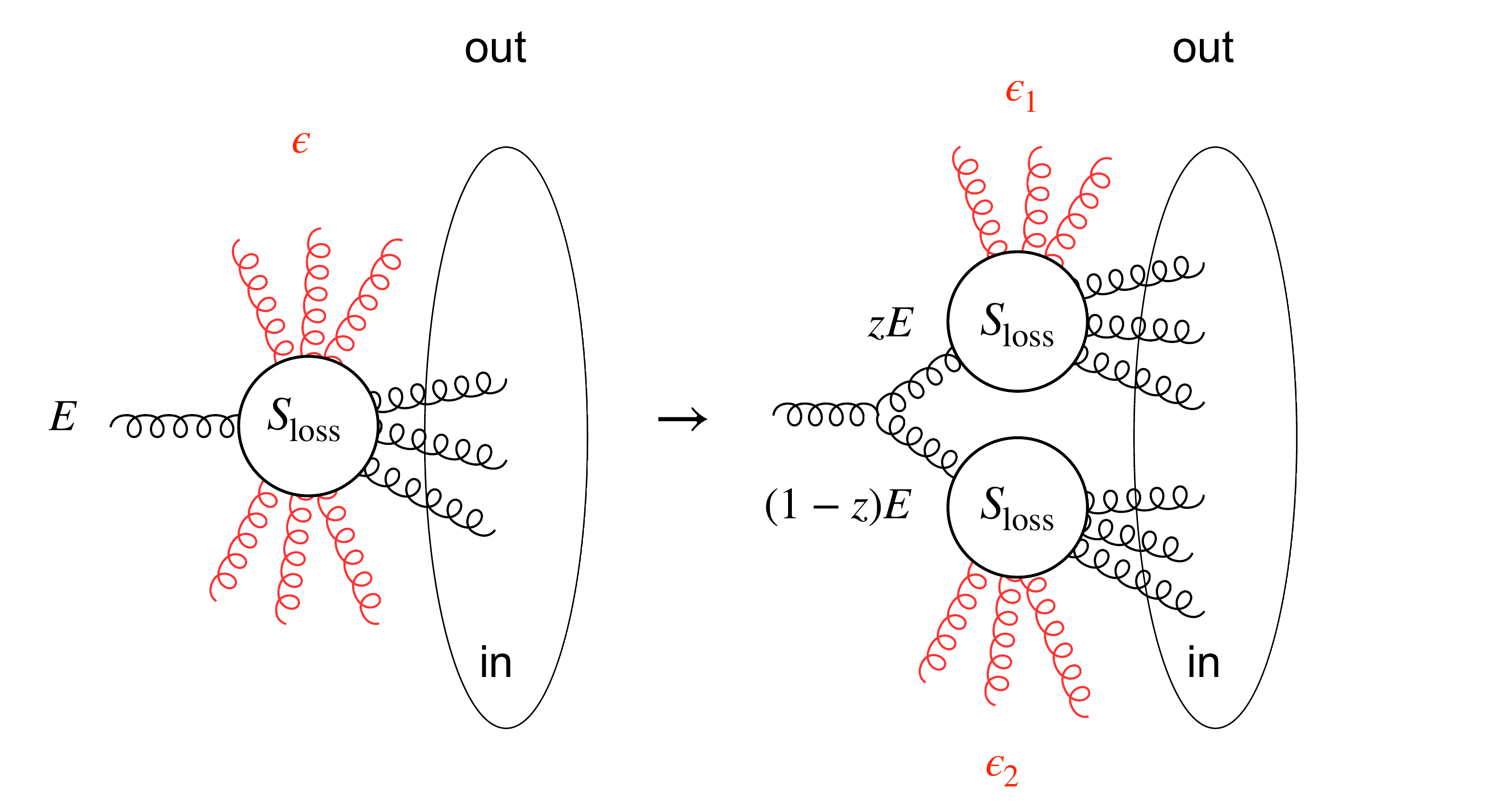}\caption{Illustration of \eqn{eq:non-lin-eq}. The left figure depicts the energy loss distribution from the initial time $t_0 $ to the final time $ t $. Between the instants $ t_0 $ and $ t_0 + \Delta t $, the initial gluon may split into two gluons. Due to rapid color decoherence, each gluon then evolves independently from $ t_0 + \Delta t $ to $ t $, contributing additively to the total energy loss, such that $ \epsilon = \epsilon_1+ \epsilon_2 $.
}\label{fig:jet-eloss}
\end{figure}

Since each splitting creates two branches that can independently contribute to the total energy loss, the evolution equation for the energy distribution outside the cone becomes non-linear, as depicted in Fig.~\ref{fig:jet-eloss}, and takes on a familiar form
\begin{align}\label{eq:non-lin-eq}
& S_{\rm loss}(\epsilon, t, t_0 ;E,\btheta)  = S_{\rm el}(\epsilon, t, t_0;E,\btheta)  \nn
&+ \alpha_s \int_{t_0}^t \rmd t_1   \int_0^1 \rmd z\, \cK(z,E)  \, \int \frac{\rmd^2 \btheta'}{(2\pi)^2 E^2}\, \cP(E(\btheta'-\btheta),t_1,t_0) \nn
& \times\left[ \int_{\epsilon_1,\epsilon_2}\,S_{\rm loss}(\epsilon_1, t, t_1 ;zE,\btheta') \,S_{\rm loss}(\epsilon_2, t, t_1 ;(1-z)E,\btheta') \, \delta(\epsilon-\epsilon_1-\epsilon_2) - S_{\rm loss}(\epsilon, t, t_1 ;E,\btheta')  \right]  \,,
\end{align}
where $\alpha_s=g^2/4\pi$ is the coupling constant. 

The evolution halts at $t=L$, after which the produced gluons propagate in vacuum, resulting in no further medium-induced effects.

Note that when $E=0$, we deduce from \eqn{eq:S-elastic} that
\beq\label{eq:null-eloss}
S_{\rm loss}(\epsilon,E=0)= \delta(\epsilon)\,,
\eeq
which is just a statement that if the parent gluon energy vanishes so does its energy loss, hence, $\epsilon=0$. This property remains true in the presence of the medium-induced cascade. It is also crucial to ensure that, even though the splitting kernel is singular in the limit $z\to 1$ and $z\to 0$ the equation is finite. This can be checked by noting that the non-linear term in \eqn{eq:non-lin-eq} admits the limit
\beq
\lim_{z\to 0}S_{\rm loss}(\epsilon_1,zE) \, S_{\rm loss}(\epsilon_2(1-z),zE) =  \delta( \epsilon_1)\, S_{\rm loss}(\epsilon_2,E)\,.
\eeq
An analogous identity applies in the limit $z\to1$, precisely eliminating the linear term.
\subsection{Evolution of quenching weights in Laplace space}
\eqn{eq:non-lin-eq}, a key result of this work, simplifies when expressed in Laplace space. To proceed, we introduce the Laplace transform of the energy loss distribution
\beq
Q_\nu (E)= \int_0^{+\infty}\rmd \epsilon \, \rme^{-\epsilon \nu}\,  S_{\rm loss}(\epsilon,E) \,.
\eeq
With this transformation, \eqn{eq:non-lin-eq} yields
\begin{align}\label{eq:non-lin-eq-LT}
& Q_\nu(t, t_0;E,\btheta)  =  Q^{\rm el}_\nu(t, t_0;E,\btheta)  + \alpha_s  \int_{t_0}^t  \rmd t_1  \int_0^1 \rmd z\, \cK(z,E)\, \int \frac{\rmd^2 \btheta'}{(2\pi)^2E^2 }\, \cP(E(\btheta'-\btheta),t_1,t_0) \nn
& \times\left[ Q_\nu( t, t_1;zE,\btheta')\,  Q_\nu( t, t_1;(1-z)E,\btheta') -  Q_\nu( t, t_1;E,\btheta')  \right]  \,.
\end{align}
and \eqn{eq:S-elastic} yields
\begin{align}
Q_\nu^{\rm el}(t, t_0;E,\btheta)  &= \int \frac{\rmd^2 \btheta'}{(2\pi)^2E^2}\, \cP(E(\btheta'-\btheta),t,t_0) \left[ \Theta(|\btheta'|<R)\, +\Theta(|\btheta'|>R)  \,\rme^{-\nu E} \right]\,. \nn
&=\rme^{-\nu E}+  \int \frac{\rmd^2 \btheta'}{(2\pi)^2E^2}\, \cP(E(\btheta'-\btheta),t,t_0) \,  \Theta(|\btheta'|<R)\, (1-\rme^{-\nu E}) \,. \nn
\end{align}
The normalization condition \eqn{eq:norm} implies
\beq
Q_0^{\rm el}(t, t_0;E,\btheta) =1\,.
\eeq
We also wish to verify that if the `in' region vanishes, then $Q_\nu(t, t_0, E) = \rme^{-\nu E}$. Conversely, if the `out' region vanishes, we must have $Q_\nu(t, t_0) = 1$. Additionally, we note that the evolution equation \eqn{eq:non-lin-eq-LT} admits a thermal fixed point:
\beq Q_\nu^{\rm th}(E) = \rme^{-\nu E}\,,\label{eq:thermal-fp}
\eeq
which corresponds to the complete decay of the parent gluon. Of course, this asymptotic fixed point is not strictly physical, as some residual energy will always remain within the `in' region after the gluon has dissipated its energy into the plasma. Nevertheless, we can interpret this by introducing an IR cutoff scale, $\omega_{\rm IR} \gtrsim T$, on the order of a few times the plasma temperature, such that only modes above this threshold are considered part of the jet.

\eqn{eq:non-lin-eq-LT} takes the same form as the collimator function, which describes the energy loss associated with the collinear vacuum cascade (cf. \eqn{eq:non-lin-eq-coll-2}) \cite{Mehtar-Tani:2017web,Mehtar-Tani:2021fud}. In Section \ref{sec:vacuum}, we will discuss how the collimator function can be naturally extended to incorporate the medium-induced cascade.

\subsection{Single parton energy loss}

The most physically relevant case for phenomenology is when a produced jet emerges out of the plasma with sufficiently high $p_T$ to be reconstructed and identified as a jet originating from a hard event. In this scenario, the energy loss that the jet experiences in the plasma is small compared to its energy, namely, $\epsilon\sim E_{\rm med} \ll E$, where $E_{\rm med}\sim \alpha^2 \hat q L^2$ is the typical gluon energy below which the gluon is fully quenched \cite{Baier:2001yt,Blaizot:2013hx,Mehtar-Tani:2017web}. Moreover,   the parent parton's angular broadening is negligible compared to the jet cone size which allows us to approximate the first term in \eqn{eq:non-lin-eq-LT} as
\begin{align}
Q^{\rm el}_\nu(t, 0 ; E,\btheta=0)& =1 \,.
\end{align}
where we have used $\cP(\q-\k,t,t_0) \approx (2\pi)^2\delta^{(2)}(\q-\k)$.  Since the energy flow is dominated by the leading parton, it approximately defines the jet center around which the jet area $|\btheta| < R$ is determined.

The same approximation is applied to the second term in \eqn{eq:non-lin-eq-LT} and we distinguish the leading (hard) branch contribution from a subleading branch for which $z\ll 1$ (or $z\ll 0$ by symmetry of the gg splitting function).  In addition, we need to approximate the splitting function by its soft limit, namely the soft radiation rate. This allows us to factorize the hard part  to obtain
\begin{align}\label{eq:non-lin-eq-jet}
& Q^{\rm hard}_\nu(t,t_0;E)  = 1 + \int_{t_0}^t  \rmd t_1  \int_0^E \rmd \omega \frac{\rmd I}{\rmd \omega}\,\left[ Q^{\rm soft}_\nu( t,t_1, \omega) -  1  \right] Q^{\rm hard}_\nu( t,t_1;E) \,,
\end{align}
where we have introduced the LO soft radiation rate (cf.~\eqn{eq:rad-spect})
\beq
\omega \frac{\rmd I}{\rmd \omega} = 2 \lim_{z\to 0 } z\cK(z,E) = \frac{\alpha_s N_c}{\pi} \sqrt{\frac{\hat q}{\omega}}\,,
\eeq
with $\omega=zE$ and $Q^{\rm hard}_\nu(t,t_1;E) \equiv Q^{\rm hard}_\nu(t,t_1,E;0)$. To keep the discussion simple and focused on the core properties of our evolution, we have intentionally neglected the fact that the spectrum changes for $\omega \gtrsim \omega_c$, where finite-size effects become significant as $t_f \sim L$. However, we emphasize that our formulation remains valid over the entire range of gluon frequencies. Furthermore, quarks can be easily incorporated into this framework.

Assuming weak quenching, i.e., $Q^{\rm hard}_\nu(t,t_0;E)\sim 1$, to first non-trivial order \eqn{eq:non-lin-eq-jet} yields
\begin{align}
& Q^{\rm hard}_\nu(t,t_0;E)  = 1 + \int_{t_0}^t  \rmd t_1  \int_0^E \rmd \omega \frac{\rmd I}{\rmd \omega}\,\left[ Q^{\rm soft}_\nu( t,t_1, \omega) -  1  \right]+\cO((1-Q^{\rm hard})^2) \,,
\end{align}

\eqn{eq:non-lin-eq-jet}  can be readily solved, yielding
\begin{align}\label{eq:lin-eq-jet-sol}
& Q^{\rm hard}_\nu(t,t_0;E)  = \exp\left\{ -  \int_{t_0}^t  \rmd t_1  \int_0^E \rmd \omega \frac{\rmd I}{\rmd \omega}\,\left[ 1-Q^{\rm soft}_\nu( t,t_1;\omega)   \right] \right\} \,.
\end{align}
where $Q^{\rm soft}$ obeys the full non-linear equation \eqn{eq:non-lin-eq-LT}. This formula is valid so long as $E\gg \alpha_s^2 \omega_c$. In the Section.~\ref{sec:steep-spect} we shall discuss this approximation in the context of the fully developed DGLAP shower.

The inverse Laplace transform leads to the energy distribution 
\begin{align}
&S^{\rm hard}(\epsilon, t,t_0;E) = \exp\left[-\int_{t_0}^t \rmd t' \int \rmd \omega \frac{\rmd I}{\rmd \omega \rmd t'}\right]  \nn
&\times\left\{\delta(\epsilon)+  \sum_{n=1}^\infty \frac{1}{n!} \prod_{i=1}^n \left[\int_{t_0}^t\rmd t_i \int\rmd \epsilon_i\,\rmd \omega_i\frac{\rmd I}{\rmd \omega_i\rmd t_i}\,  S^{\rm soft }(\epsilon_i, t,t_i;\omega_i) \right]\delta(\epsilon-\sum_{i=1}^n\epsilon_i)\right\}\,,
\end{align}
which is a  generalization of the Poisson-like energy loss distribution first proposed in \cite{Baier:2001yt} that can be recovered by neglecting the soft gluon shower, or by setting $R = 0$, .i.e., 
\beq 
\lim_{R\to 0}S^{\rm soft }(\epsilon_i, t,t_i;\omega_i)  \to \delta(\epsilon_i-\omega_i)\,.
\eeq
or equivalently, $\lim\limits_{R\to 0}Q^{\rm soft}_\nu( t,t_1; \omega) =\rme^{-\nu \omega}\,$. 
\subsection{Perturbation theory around the BDMS approach }

In order to gain insight on the analytic properties of \eqn{eq:lin-eq-jet-sol} we can look for a solution near the Baier-Dokshitzer-Mueller-Schiff (BDMS) one \cite{Baier:2001yt}, that is,
\beq\label{eq:linear-Q}
Q^{\rm soft}_\nu( t,t_1;\omega) = \rme^{-\nu \omega} \left[1 + q_\nu( t,t_1;\omega) \right]\,,
\eeq
where we shall assume $q \ll 1$. Inserting  \eqn{eq:linear-Q} in \eqn{eq:non-lin-eq-LT} keeping only the linear terms in $q(\omega)$ we write
\begin{align}\label{eq:lin-eq}
& q_\nu(t,t_0;\omega,\btheta)  = \int \frac{\rmd^2 \btheta'}{(2\pi)^2\omega^2}\, \cP(\omega(\btheta'-\btheta),t,t_0)  \Theta(|\btheta'|<R) \left( \rme^{\nu \omega}+1\right)\nn
&+ \alpha_s \int_{t_0}^t \rmd t_1  \int_0^1 \rmd z \cK(z)  \, \int \frac{\rmd^2 \btheta'}{(2\pi)^2\omega^2}\, \cP(\omega(\btheta'-\btheta),t_1,t_0) \nn
& \times\left[ 2 q_\nu( t, t_1; z\omega, \btheta')  -  q_\nu(t,t_1; \omega,\btheta')  \right]  \,,
\end{align}
whose solution can be expressed as
\beq
q_\nu(t,t_0;\omega,\btheta) =   \int \frac{\rmd^2 \btheta'}{(2\pi)^2\omega^2}\,   \int_0^1 \rmd x\, \Theta(|\btheta'|<R) \left( \rme^{\nu x \omega}+1\right) D(t, x\omega,\btheta'|t_0,\omega,\btheta)\,,
\eeq
where $D(t, x\omega,\btheta'|t_0,\omega,\btheta)\simeq D(x,\btheta'-\btheta,\omega,t-t_0)$, is the inclusive distribution measuring a gluon of momentum $x\omega$ at the angle $\btheta'$ at time $t$ in a parent gluon of frequency $\omega$, which was discussed extensively in \cite{Blaizot:2014rla}.  It obeys the integral equation
\begin{align}\label{eq:lin-eq-incl}
& D(t, x\omega,\btheta'|t_0,\omega,\btheta)\  =  \cP(\omega(\btheta'-\btheta),t,t_0) \delta(1-x)\nn
&+ \alpha_s \int_{t_0}^t \rmd t_1  \int_0^1 \rmd z \cK(z,\omega)  \, \int \frac{\rmd^2 \btheta'}{(2\pi)^2\omega^2}\, \cP(\omega(\btheta'-\btheta),t_1,t_0) \nn
& \times\left[ 2 D(t, xz\omega,\btheta'|t_1,z\omega,\btheta)  -   D(t, x\omega,\btheta'|t_1,\omega,\btheta)\right]  \,.
\end{align}
Apart from the first elastic term, we recognize the evolution equation for the inclusive gluon distribution \cite{Blaizot:2013vha,Blaizot:2014rla}. To make the connection explicit we apply the transverse-broadening operator $\frac{\del}{\del t_0 }  -  \int_\bell \cC(\bell) \otimes$ to both sides of the equation which allow us to use the identity
\beq
\frac{\del}{\del t_0 }  \cP(\omega(\btheta'-\btheta),t,t_0)  + \int_\bell \cC(\bell) \otimes \cP(\omega(\btheta'-\btheta)-\bell,t,t_0) =0\,,
\eeq
This manipulation turns our integral equation into an integro-differential one. As a result, the elastic term cancels out and  the $\cP$ factor in the second term collapses since the time derivative acting on the lower limit of $t_1$ sets $t_1=t_0$ in $\cP$ yielding
\begin{align}\label{eq:lin-eq-2}
& \frac{\del }{\del t_0}q(t-t_0|\k,\omega)  =- \int_\bell \cC(\bell) \otimes q(t-t_0|\k-\bell,\omega) \nn
&+   \alpha_s  \int_0^1 \rmd z \cK(z,\omega)  \left[ q( t- t_1|z\q,z\omega) + q(t-t_1 |(1-z)\q,(1-z)\omega)  -  q(t-t_1  |\q,\omega)  \right]  \,.
\end{align}
which must be solved with the initial condition
\beq
q(t=0|\k,\omega) =  \Theta(|\k|<RE) \left( \rme^{\nu E}+1\right)\,.
\eeq
Physically, this result indicates that the residual energy within the jet is determined by the average distribution of shower gluons confined to the jet region. We shall leave a detailed study of this equation in the context of energy loss to a future work.

\section{Non-linear DGLAP evolution }\label{sec:vacuum}

\subsection{Coherent vacuum parton shower }
Up to this point, we have neglected the collinear parton cascade that arises from the decay of the initial parton's virtuality, which is responsible for jet fragmentation in vacuum. This cascade occurs prior to the medium-induced cascade, making the solution to \eqn{eq:non-lin-eq} the initial condition for the early-time non-linear DGLAP evolution.

Following the same construction as for the medium-induced cascade, we write

\begin{align}\label{eq:non-lin-eq-coh}
&S_{\rm loss}(\epsilon,E,\k,\theta_{\rm max})  = S_{\rm loss}^{\rm med}(\epsilon,E,\k,R)  \nn
&+ \frac{\alpha_s}{2\pi^2}\int_0^{\theta_{\rm max}} \rmd \theta\int \frac{\rmd^2 \q}{\q^2}  \int_0^1 \rmd z p_{gg}(z)  \delta\left(\theta-\frac{|\q|}{z(1-z)}\right) \Theta_{\rm PS}(\theta,z) \nn
& \times\left[ \int_{\epsilon_1,\epsilon_2} S_{\rm loss}(\epsilon_1, zE,\q+z\k, \theta)    S_{\rm loss}(\epsilon_2,(1-z)E,-\q+(1-z)\k,\theta)   \delta(\epsilon-\epsilon_1-\epsilon_2)\right. \nn
&\left.-  S_{\rm loss}(\epsilon_1, E,\k,\theta)   \right]  \,,\,
\end{align}
where the non-linear term describes the splitting of a gluon with transverse momentum $\k$ into two gluons of momenta $-\q+z\k$ and $\q+(1-z)\k$, respectively \cite{Catani:1990rr}. The initial condition is $ S_{\rm loss}^{\rm med}$ encodes medium-induced gluon cascade and is solution of the master equation \eqn{eq:non-lin-eq}. The maximum angle $ \theta_{\rm max} $ serves as the evolution variable, running up to angles of order $R$, and is related to the virtuality of the decaying parton $\mu $ through $ \mu \sim E \theta_{\rm max} $. The small $R$ approximation will lead to a further factorization.  The angular constraint $\theta< \theta_{\rm max}$
arises from the leading infrared (IR) contributions in perturbative QCD. To correctly account for leading soft and collinear logarithms, the kinematically available phase space for parton emission must be restricted to the angular-ordered region, where the branching angles decrease progressively along the cascade from the hard vertex to the final state \cite{Mueller:1981ex,Bassetto:1982ma,Dokshitzer:1982ia}. Outside this angular-ordered region, different parton emitters act coherently, resulting in destructive interference, where the azimuthally integrated distribution vanishes at leading order. This phenomenon is known as color coherence \cite{Dokshitzer:1991wu}.

The angular ordered constraint $ \theta < \theta_{\rm max}$ ensures the proper accounting of leading soft and collinear logarithms and is a result of color coherence \cite{Dokshitzer:1991wu}. The kinematically available phase space for parton emission is limited to the angular-ordered region, where branching angles decrease sequentially along the cascade from the hard vertex to the final state \cite{Mueller:1981ex,Bassetto:1982ma,Dokshitzer:1982ia}. Destructive interference suppresses large-angle emissions outside this region, as multiple partons behave coherently as a single effective emitter.

Interaction with the medium color charges tends to alter the color coherence of the jet, thereby opening up the phase space for radiation outside the angular-ordered region. This anti-angular-ordered radiation \cite{Mehtar-Tani:2010ebp,Mehtar-Tani:2011hma}, occurs at medium scales and does not generate large collinear logarithms, such as $\ln (R/\theta_c)$. However, it can be enhanced by the length of the medium \cite{Mehtar-Tani:2017ypq}. Color decoherence radiation takes place when the medium resolves individual color charges within the jet.

In the simplest case of an antenna forming an angle $\theta $, if the medium's resolution scale, which is related to transverse momentum broadening as $(\hat{q} L)^{-1/2} $, is smaller than the typical in-medium antenna size, $ r_\perp \sim \theta L $, i.e., $ (\hat{q} L)^{-1/2} < \theta L $, the two emitters radiate independently \cite{Casalderrey-Solana:2011ule,Mehtar-Tani:2012mfa,Mehtar-Tani:2017ypq}. This inequality implies that for incoherent radiation to occur, the angle between the hard collinear partons must satisfy the condition
\beq
\theta > \theta_c \equiv \frac{1}{(\hat{q} L^3)^{1/2}}\,,
\eeq
where $ \theta_c $ is the in-medium coherence angle \cite{Mehtar-Tani:2011hma,Mehtar-Tani:2012mfa,Mehtar-Tani:2017ypq}, which also corresponds to the minimum angle for medium-induced radiation \cite{Baier:1996kr}. Collinear splittings at smaller angles act coherently as a single color charge, and therefore do not impact the observable.

Apart from the angular constraint, the transverse momentum of the collinear splitting must exceed the typical transverse momentum for medium-induced splittings, given by $ k_\perp \equiv \sqrt{z(1-z) \hat{q}} $ (see \cite{Blaizot:2015lma} and references therein).

\begin{figure}
\begin{center}
\includegraphics[width=0.6\textwidth]{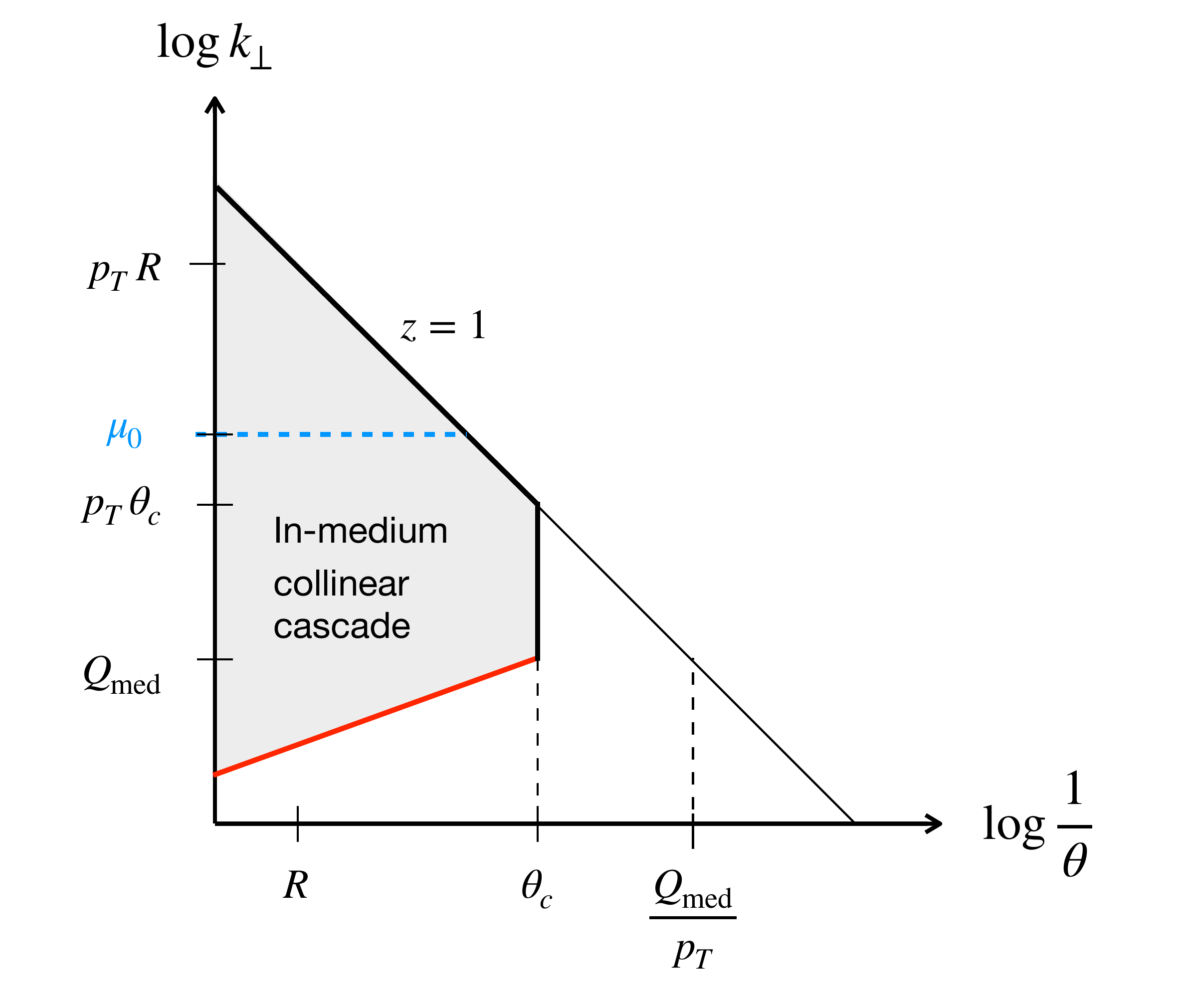}
\end{center}
\caption{ Lund-plane representation of the phase-space for in-medium vacuum shower. The dynamics near the boundaries $\theta=\theta_c$ and $k_\perp=(\omega \hat q)^{1/4}$, which correspond to the medium resolution angle and the medium-induced transverse momentum scale is driven by the details for the interactions with the medium. However, the physics in the hard sector, namely, the upper corner of the triangle is fully determined by DGLAP vacuum physics. The large separation of scales is illustrated by the intermediate factorization scale $\mu_0$. The medium  scale is typically given by $Q_{\rm med}\sim \sqrt{\hat q L}$. }
\label{fig:Lund-Plane}
\end{figure}

The angular and transverse momentum constraints described above define the phase space for resolved vacuum emissions, as illustrated in Fig.~\ref{fig:Lund-Plane}, and can be expressed as
\beq\label{eq:PS}
\Theta_{\rm PS}(\theta,z)  = \Theta(\theta>\theta_c) \Theta(|\q|> (z(1-z) E\hat q )^{1/4}) \,.
\eeq
Collinear splittings outside this phase space are either unresolved by the medium or occur outside the medium, and therefore do not contribute to inclusive jet production.

\subsection{Non-linear evolution for $\texorpdfstring{\ln (R/\theta_c)}{\ln (R/\theta_c)}$ resummation}

Again, it is convenient to express \eqn{eq:non-lin-eq-coh} equation in Laplace space
\begin{align}\label{eq:non-lin-eq-coll}
& Q_\nu(E,\k,\theta_{\rm max})  = Q_\nu^{\rm med}(E,\k)  + \frac{\alpha_s}{2\pi^2}\int_0^{\theta_{\rm max}}\rmd \theta\int \frac{\rmd^2 \q}{\q^2}  \int_0^1 \rmd z \,p_{gg}(z)  \delta\left(\theta-\frac{|\q|}{z(1-z)}\right)\nn
& \times\left[ Q(zE,\q+z\k,\theta)    Q_\nu((1-z)E,-\q+(1-z)\k,\theta)  -  Q_\nu(E,\k,\theta)   \right]  \,,
\end{align}
To leading logarithmic accuracy, $\theta\ll \theta_{\rm max} \sim R$, therefore we may neglect variation of energy loss due to small deviations of collinear parton angles, i.e., $Q_\nu(E,\k,R)  \simeq Q_\nu(E,\k=0,R)\equiv Q_\nu(E,R)$, that are power suppressed. Upon this simplification and after integration over $\q$ then taking the derivative w.r.t. $\theta_{\rm max} \sim R$ \footnote{Throughout the evolution, we distinguish between $R$ and $\theta_{\rm max} $, the actual evolution variable, as the initial condition also depends on $R$. We set $\theta_{\rm max}=R$ only at the final stage.}, we recover the renormalization group equation (RG) for the quenching weight from \cite{Mehtar-Tani:2017web,Mehtar-Tani:2021fud}
\begin{align}\label{eq:non-lin-eq-coll-2}
& \theta\frac{\del }{\del \theta}Q_\nu(E,\theta)  =   \frac{\alpha_s}{\pi} \int_0^1 \rmd z \,  p_{gg}(z) \, \left[ Q_\nu(zE,\theta)    Q_\nu((1-z)E,\theta)  -  Q_\nu(E,\theta)   \right]  \,.
\end{align}
For the case of a running coupling, we should use $\alpha_s\equiv \alpha_s(z(1-z)\theta))$.  Here, we can ignore the phase space constraint since the equation is IR finite, i.e., when  $z\to 0$ or $z\to 1$,  and so long as $\theta>\theta_c$.  Indeed, the $z$ integral is cutoff dynamically in the hard collinear sector. A similar non-linear DGLAP evolution has been derived for track functions \cite{Chen:2022muj}, which measure a subset of hadrons, such as charged particles, within the jet \cite{Chang:2013rca}. The connection between the non-linear evolution of energy loss and track functions was noted recently \cite{Barata:2024bmx}. However, this straightforward extension of the track function approach to jet energy loss incorrectly assumes that the jet energy loss distribution is boost invariant. As discussed in this paper, the energy loss distribution depends not only on the energy fraction $ x $ and the factorization scale $ \mu $, but also on the initial energy $E $, which has motivated our (local) formulation in Laplace space.

This non-linear DGLAP equation is expressed as an evolution in angles to account for angular ordering along the cascade \cite{Dokshitzer:1991wu,Dasgupta:2014yra}. We may also write a virtuality evolution, defining $\mu\equiv ER$ and $Q_\nu(E,R)\to Q_\nu(E,\mu)$, we obtain
\begin{align}\label{eq:non-lin-eq-coll-mu}
& \mu \frac{\del }{\del \mu}Q_\nu(E,\mu)  =  \frac{\alpha_s}{\pi} \int_0^1 \rmd z \,p_{gg}(z) \, \left[ Q_\nu(zE,z\mu)    Q_\nu((1-z)E,(1-z)\mu)  -  Q_\nu(E,\mu)   \right]  \,.
\end{align}
Neglecting angular ordering amounts to removing the factors $z$ and $(1-z)$ multiplying $\mu$ in the r.h.s. This is justified so long as the $z$ integral does not receive large contributions from the soft sector. 

This leading logarithmic (LL) equation resums the large collinear logs: $\ln (R/\theta_c)$.  

\subsection{Linear evolution for $\texorpdfstring{\protect\ln (1/R)}{\ln (1/R)}$ resummation }
While the discussion above focused on collinear splittings within the jet ($ \theta < R $), \eqn{eq:non-lin-eq-coll-mu} remains applicable at larger angles to leading logarithmic (LL) accuracy. For $\theta \gg R $, one of the branches lies outside the jet cone, and, assuming negligible radiation back into the jet, all of its energy is lost, leading to $Q_\nu = 1 $. In this limit, the non-linear equation \eqn{eq:non-lin-eq-coll-mu} simplifies to a linear DGLAP-like equation. Thus, \eqn{eq:non-lin-eq-coll-mu} also resums large $ \ln R $ terms, corresponding to early out-of-cone vacuum radiation. In this case, we obtain
\beq
Q_\nu(zE,\theta>R) \simeq\rme^{-zE \nu} \,\quad \text{and} \quad  Q_\nu((1-z)zE,\theta>R) \simeq\rme^{-(1-z)E \nu}\,,
\eeq
for the $z$ and $1-z$ branches, respectively.  Hence, when  $ \theta \gg R$, or $\mu >Q\equiv  ER$, we obtain
\begin{align}\label{eq:lin-eq-coll}
& \mu\frac{\del }{\del {\mu}}Q_\nu(E,\mu) =\frac{\alpha_s}{\pi} \int_0^1 \rmd z \,  p_{gg}(z) \nn
&  \,\times \left[  \rme^{-(1-z) zE \nu}Q_\nu(zE,\mu)    +  \rme^{-zE \nu}Q_\nu((1-z)E,\mu)  -  Q_\nu(E,\mu)   \right]  \,,
\end{align}

Recalling the relationship with the jet function \eqn{eq:jet-function} to derive the corresponding evolution equation, we find, upon performing the inverse Laplace Transform of the first term
\begin{align}\label{eq:J-term-1}
\mu\frac{\del }{\del {\mu}}J(x,E,\mu) &=\frac{\alpha_s}{\pi} \int_0^1 \rmd z \,  p_{gg}(z) \int \frac{\rmd \nu}{2\pi i } \rme^{(1-x) p_T \nu - (1-z)E\nu} Q_{x\nu}\left(\frac{z E}{x},\mu\right)\nn
&=\frac{\alpha_s}{\pi} \int_0^1 \frac{\rmd z}{z} \,  p_{gg}(z) \int \frac{\rmd \nu'}{2\pi i } \rme^{(1-x) E \nu'} Q_{(x/z)\nu}\left(\frac{z E}{x},\mu\right)\nn
&=\frac{\alpha_s}{\pi} \int_0^1 \frac{\rmd z}{z} \,  p_{gg}(z) J\left(\frac{x}{z},zE,\mu\right)\,,
\end{align}
where in the second to last step we have changed variables, namely,  $\nu'=z\nu$.

Finally, using the symmetry $z \leftrightarrow 1-z$ of the equation and replacing the splitting function by the regularized Altarelli-Parisi splitting function \cite{Altarelli:1977zs}
\beq\label{eq:AP}
P_{gg}(z) =p_{gg}(z)_++p_{gg}(1-z) \,,
\eeq
we finally obtain the DGLAP-like equation
\begin{align}\label{eq:J-DGLAP-1}
& \mu\frac{\del }{\del {\mu}}J(x,E,\mu) =\frac{\alpha_s}{\pi} \int_0^1 \frac{\rmd z}{z} \,  P_{gg}(z)  J(x/z,zE,\mu) \,.
\end{align}
In the pure vacuum case there is no other scale other than $E$ and $\mu$, hence, owing to boost invariance we can combine the two scales to form dimensionless arguments, i.e., $J(x,E,R,\mu)\simeq J(x,\mu/ER)$, where we have restored  the implicit $R$ dependence. However, the presence of medium scales breaks this property and thus and extra $E$ dependence remains, as discussed in Section.~\ref{sec:eloss-dist}.

Interestingly, with this observation, \eqn{eq:J-DGLAP-1} reduces to the recently proposed form of the jet function evolution equation valid at NLL \cite{Lee:2024tzc,vanBeekveld:2024jnx},
\begin{align}\label{eq:J-DGLAP-2}
& \mu\frac{\del }{\del {\mu}}J\left(x,\frac{\mu}{ER}\right) =\frac{\alpha_s}{\pi} \int_0^1 \frac{\rmd z}{z} \,  P_{gg}(z) \, J\left(\frac{x}{z},\frac{\mu}{zER}\right)  \,.
\end{align}
Extending beyond leading logarithmic accuracy requires matching coefficients at the transition scale $\mu = ER$, which connects the linear DGLAP-like renormalization group evolution for $\mu \gg ER$ with the non-linear evolution for $ \mu \ll ER$ \cite{Mehtar-Tani:2024smp}.

This leading logarithmic approach captures the essential features of the all-order structure of our formula at leading power in $R^2$, while retaining certain $R^2$ contributions associated with energy loss. 

At higher orders, it is necessary to compute matching coefficients that account for splittings at angles around $\theta \sim R$. This is discussed in Section~\ref{sec:higher-orders}. A factorization formula for this case was recently derived in \cite{Mehtar-Tani:2024smp}, specifically for the regime $R \sim \theta_c$, enabling a detailed treatment of the infrared region in jet evolution. In the regime $R \gg \theta_c$, collinear logarithms $\ln(R/\theta_c)$ can be resummed, as discussed above, where interference contributions are suppressed. Additional matching is required at angles $\theta \sim \theta_c$, where the factorization framework developed in \cite{Mehtar-Tani:2024smp} is applicable.

\subsection{Strong quenching limit }

In general, the non-linear equations derived in the previous sections cannot be solved analytically. However, it is possible to linearize the equation in the two limiting cases of strong and weak quenching, respectively. In the following, we will focus on the former case.

In the strong quenching limit one can expend around the thermal fixed point, \eqn{eq:thermal-fp} that corresponds to ``total'' energy loss:
\beq
Q_\nu(E,\mu) = \rme^{-\nu E} (1+ q_\nu(E,\mu))\,,
\eeq
where we assume $q_\nu(E,\mu)\ll 1$.

To linear order in $q$, \eqn{eq:non-lin-eq-coll-mu} yields
\begin{align}\label{eq:strong-q}
\mu\frac{\del }{\del {\mu}}q_\nu(E,\mu) &\simeq \frac{\alpha_s}{\pi} \int_0^1 \rmd z \,  p_{gg}(z) \left[ q_\nu(zE,\mu)  +  q_\nu((1-z) E,\mu) - q_\nu(E,\mu)\right] \,.\nn
\end{align}
We can easily check that the following integral of $q$ over $E$ is an RG invariant:
\beq
\mu\frac{\del }{\del {\mu}} \int_0^\infty \frac{\rmd E}{E^2}\,q_\nu(E,\mu)  =0\,.
\eeq
Using the $+$ prescription and \eqn{eq:AP} we obtain
\begin{align}\label{eq:strong-q-2}
\mu\frac{\del }{\del {\mu}}q_\nu(E,\mu) &\simeq \frac{\alpha_s}{\pi} \int_0^1 \rmd z \,  P_{gg}(z)  \,q_\nu(zE,\mu) \,.
\end{align}
Defining the Mellin transform of $q_\nu$
\beq
q_{\nu,N}(\mu) =\int_0^\infty \rmd E E^{-N} \,q_\nu(E,\mu)\,,
\eeq
and its inverse
\beq
q_\nu(E,\mu) =\int \frac{\rmd N}{2\pi i } \,E^{N-1} \,q_{\nu,N}(\mu)\,,
\eeq
\eqn{eq:strong-q-2} yields
\begin{align}\label{eq:strong-q-3}
\mu\frac{\del }{\del {\mu}}q_{\nu,N}(\mu)  &\simeq  \gamma(N)\, q_{\nu,N}(\mu) \,.
\end{align}
where $\gamma(N)$ is the DGLAP anomalous dimension
\beq
\gamma(N)  = \int_0^1 \rmd z \, z^{N-1} \, P_{gg}(z) \,.
\eeq
The solution can be written readily as
\beq
q_\nu(E,\mu) =  \int_{c-i\infty}^{c+i\infty}\frac{\rmd N}{2\pi i } E^{N-1} \exp\left[ \frac{\alpha_s}{\pi} \gamma(N) \ln \frac{\mu}{\mu_0} \right] q_{\nu,N}(\mu_0)\,.
\eeq
This is the general solution of the RG evolution down to some medium scale. One very important consequence is this limit is the cancellation of the $ \mu\sim ER$ dependence due to vacuum evolution and its replacement by $E\theta_c$.

The solution for $Q$ reads
\begin{align}\label{eq:sol-strong-Q}
Q_\nu(E,\mu) &=   \rme^{-\nu E} +  \rme^{-\nu E} \int\frac{\rmd N}{2\pi i } \,E^{N-1}\,\rme^{ \frac{\alpha_s}{\pi} \gamma(N) \ln \frac{\mu}{\mu_0} } \int_0^\infty \rmd E' E'^{-N} q_\nu(E',\mu_0)\, \,.\nn
&=\rme^{-\nu E} +  \rme^{-\nu E} \int_0^1\rmd z \,q_\nu(zE,\mu_0)  \int\frac{\rmd N}{2\pi i } \,z^{-N}\,\rme^{ \frac{\alpha_s}{\pi} \gamma(N) \ln \frac{\mu}{\mu_0} }  \,.\nn
\end{align}
Since $q_\nu$ has only support for $E'<E$, we have $z<1$.

At leading logarithm accuracy, we can choose $\mu_0 \simeq E \theta_c$, and use the solution of \eqn{eq:lin-eq} as an initial condition for $q_{(\nu,n)}(E,\mu)$.
In frequency space \eqn{eq:sol-strong-Q} becomes
\begin{align}\label{eq:sol-strong-S}
S_{\rm loss}(\epsilon,E,\mu)
&=\delta(\epsilon-E)+   \int_0^1\rmd z \, \left[S_{\rm loss}(\epsilon,zE,\mu_0) -\delta(\epsilon-E)\right]  \int\frac{\rmd N}{2\pi i } \,z^{-N}\,\rme^{ \frac{\alpha_s}{\pi} \gamma(N) \ln \frac{\mu}{\mu_0} }  \,.\nn
\end{align}
Which can simplified further in the large $N$, corresponding to $z\sim1$, limit. We shall discuss the relevance of this approximation in the next section. 

A thorough investigation of this limit involves solving for the initial conditions, which can be done analytically with additional simplifying assumptions. We defer this task to future work to maintain focus on the broader structure of our framework. Ultimately, we aim to solve  the non-linear equations numerically for phenomenological applications.

\section{Sudakov suppression of jets }\label{sec:steep-spect}

\subsection{The spectrum's power-law index as an expansion parameter}

The power exponent of the spectrum has long been identified as a important amplifying effect in jet energy loss \cite{Baier:2001yt}. Indeed, assuming a constant shift $\epsilon \ll p_T$ of the jet spectrum due to energy loss and expanding to second order we have
\beq
\frac{\rmd \sigma}{\rmd p_T} \sim \frac{1}{(p_T+\epsilon)^n} \simeq \frac{1}{p_T^n} \left(1-\frac{n \epsilon}{p_T} +\cO(n \epsilon^2)\right)\,,
\eeq
where we see that a small shift $\epsilon \sim p_T/n \ll p_T$ can have order one effect on the jet spectrum in heavy ion collisions.  In terms of the jet function, assuming $H(E=p_T/x)\sim (x/ p_T)^n$, we can approximate \eqn{eq:factorization}
\beq\label{eq:factorization-steep}
\frac{\rmd \sigma_{\rm incl}}{\rmd p_T}  \approx  H(E=p_T) \int_0^{1}\rmd x\, x^{n-1}\,J(x,E=p_T/x,R,\mu=p_T R/x) \,.
\eeq
It follows that when $n\gg 1$ the integral over $x$ receives the largest contribution close to the threshold region $x\sim 1$, with the result of inducing potentially large $\ln n$ which may provide an explanation as to the observed quantitative importance of the small-$R$ resummation even at moderate values of $R$ \cite{Dasgupta:2016bnd,Kang:2016mcy}.

In Laplace representation, from \eqn{eq:jet-function} we can rewrite \eqn{eq:factorization-steep} as
\beq\label{eq:jet-function-2}
J\left(x,\frac{p_T}{x}\right) = \frac{p_T}{x} \int \frac{\rmd \nu}{2\pi i} \, \rme^{ \frac{(1-x)}{x}p_T \nu}Q_{\nu}\left(\frac{p_T}{x}\right)  \, .
\eeq
It follows that,
\beq\label{eq:factorization-steep-LT}
\int_0^{1}\rmd x\, x^{n-1}\,J\left(x,\frac{p_T}{x}\right) =p_T \int \frac{\rmd \nu}{2\pi i} \,  \int_0^{1}\rmd x\, x^{n-2} \rme^{ \frac{(1-x)}{x} p_T \nu}Q_{\nu}\left(\frac{p_T}{x}\right)  \,.
\eeq
Inserting the solution for strong quenching given by \eqn{eq:sol-strong-Q} we obtain
\beq\label{eq:factorization-steep-LT-2}
&& \int_0^{1}\rmd x\, x^{n-1}\,J\left(x,\frac{p_T}{x}\right) =p_T \int \frac{\rmd \nu}{2\pi i} \,  \int_0^{1}\rmd x\, x^{n-2} \rme^{ \frac{(1-x)}{x} p_T \nu} \nn
&& \left[\rme^{-\nu p_T/x} +  \rme^{-\nu p_T/x} \int\frac{\rmd N}{2\pi i } \,\,\rme^{ \frac{\alpha_s}{\pi} \gamma(N) \ln \frac{\mu}{\mu_0} } \int_0^1\rmd z z^{-N} q_\nu(zp_T/x,\mu_0)\right] \,\nn
&& =\frac{1}{n-1} p_T \delta(p_T)+p_T   \,\,\rme^{ \frac{\alpha_s}{\pi} \gamma(n) \ln \frac{\mu}{\mu_0} } \int_0^1\rmd y y^{n-2} \int \frac{\rmd \nu}{2\pi i} \, \, \rme^{-\nu p_T} q_\nu(p_T/y,\mu_0) \,.\nn
\eeq
where in the last line we have integrated over $x$ and $N$.

Expression the result as function of the energy loss distribution and neglecting the irrelevant  $\delta(p_T)$ term, the  cross-section reads
\begin{align}\label{eq:factorization-steep-2}
&\frac{\rmd \sigma_{\rm incl}}{\rmd p_T}  \approx p_T \, H(E=p_T,\mu=p_TR)\,  \,\,\rme^{ \frac{\alpha_s}{\pi} \gamma(n) \ln \frac{\mu}{\mu_0} } \int_0^1\rmd y y^{n-2} \int \frac{\rmd \nu}{2\pi i} \, \,\rme^{\frac{1-y}{y}\nu p_T} Q_\nu(p_T/y,\mu_0) \,\nn
&= p_T \, H(E=p_T,\mu=p_TR)\,  \,\,\rme^{ \frac{\alpha_s}{\pi} \gamma(n) \ln \frac{\mu}{\mu_0} } \int_0^1\rmd y \, y^{n-2}  \, \, S_{\rm loss}((1-y)p_T/y, p_T/y,\mu_0) \,
\end{align}
where we have used \eqn{eq:eloss-LT}.

\subsection{The $R_{AA}$ in the double-logaritmic approximation (DLA)}

Now, changing variable to $\epsilon=(1-y)p_T$ and expanding the above expression around $n\to \infty$ limit. We have
\beq
\gamma(n)  \to  - 2N_c \left( \ln n +\gamma_E-\frac{11}{12}\right) +\cO(1/n)\,.
\eeq
with $\gamma_E \approx 0.5772$ the Euler constant, and
\beq
&& p_T \int_0^1\rmd y \, y^{n-2}  \, \, S_{\rm loss}((1-y)p_T, p_T/y) =  \int_0^{+\infty}\rmd \epsilon \, \left(\frac{p_T}{p_T+\epsilon}\right)^{n-2}  \, \, S_{\rm loss}\left(\epsilon, p_T+\epsilon\right) \nn
&&\quad \quad\quad=  \int_0^{+\infty}\rmd \epsilon \, \rme^{(n-2) \frac{\epsilon}{p_T} } \, \, S_{\rm loss}\left(\epsilon, p_T\right)  + \cO(\epsilon/p_T)+\cO(n \epsilon^2/p_T^2)\nn
&&\quad\quad\quad\simeq  Q_{\nu=p_T/n}(p_T)\,.
\eeq
As a result, we obtain for the large $n$ limit of the inclusive jet production cross-section in the double logarithmic approximation (DLA):
\begin{align}\label{eq:factorization-steep-res}
\frac{\rmd \sigma_{\rm med}}{\rmd p_T}  \approx  \, H(p_T,\mu)\,  \,\,\exp\left(- \frac{2\alpha_s N_c}{\pi} \ln n \ln \frac{\mu}{\mu_0} \right) Q_{\nu=p_T/n}(p_T,\mu_0)\ \,,
\end{align}
where we have assumed that $\mu\sim p_TR$ and $\mu_0\sim p_T \theta_c$.

This simple asymptotic result exhibits two main features of jet quenching:
\begin{enumerate}
\item In addition to the leading parton energy loss there is a Sudakov suppression due to the quenching of micro-jets within the jet. As result, if the a jet is measured it is composed of a single hard subjet of angular size $\theta_c$.
\item A second important property is the cancelation of the $R$ dependence generated by the DGLAP evolution. Of course, there subsides an $R$ dependence in the medium-induced cascade as part of the quenching factor.
\end{enumerate}
These effects were already discussed in \cite{Mehtar-Tani:2017web}.
In the leading logarithmic approximation (LLA), the vacuum cross-section can be expressed as:

\begin{align}
\frac{\rmd \sigma^{\rm vac}}{\rmd p_T} \approx H(p_T, \mu = p_T R)\,.
 \end{align}

Asymptotically, in the strong quenching limit, the nuclear modification factor is expected to behave approximately as:

\beq R_{AA} = \frac{\rmd \sigma^{\rm vac} / \rmd p_T}{\rmd \sigma^{\rm med} / \rmd p_T} \approx \rme^{-\frac{2\alpha_s N_c}{\pi} \ln(n) \ln(R/\theta_c)} \, Q_{\nu=p_T/n}(p_T, R)\,, \eeq
where $Q_{\nu=p_T/n}(p_T, R)$ encapsulates the quenching factor for a single color charge. 

In order to observe the Sudakov suppression factor in experimental data, it is necessary to impose an infrared (IR) cutoff to eliminate soft particles. These soft particles are responsible for the residual $R$-dependence in the medium-induced component, which leads to an enhancement of $R_{AA}$ due to energy recovery via soft emissions at large angles. With such a cutoff applied, we predict a systematic decrease in the nuclear modification factor $R_{AA}$ as the jet cone size $R$ increases.

\section{Non-perturbative effects and medium response }\label{sec:NP-effects}

Although our evolution equation is infrared finite the dynamics around the temperature scale should be described by kinetic theory. However, since soft modes relax quasi-instantaneously we can encode formally this physics by introducing a non-perturbative term that accounts for the thermalized energy whose distribution is known. By varying the infrared separation scale we can gauge to uncertainties associated with the non-perturbative regime. Such an approach allows us to account for the effect of medium response on energy loss. Similar strategy was previously implemented in Monte Carlos simulations such as CoLBT \cite{Chen:2017zte} and MARTINI \cite{Park:2018acg} where jet quanta that fall below an IR cutoff are modeled as sources for subsequent hydro evolution.

To illustrate our point, take for instance the expression given by \eqn{eq:lin-eq-jet-sol}. It would be modified as follows
\begin{align}\label{eq:lin-eq-jet-sol-NP}
& Q^{\rm hard}_\nu(t,t_0;E)  \to  \exp\left\{ - \int_{t_0}^t  \rmd t_1  \int_{\omega_{\rm th}}^E \rmd \omega \frac{\rmd I}{\rmd \omega \rmd t_1}\,\left[ 1-Q^{\rm soft}_\nu( t,t_1;\omega)   \right] - F_{\rm NP}(\nu, \omega_{\rm th}) \right\} \,,
\end{align}
where here, the function $ F_{\rm NP}$ account for the energy redistribution of primary gluon radiation with energy below $\omega_{\rm th}$. It reads
\beq
F_{\rm NP}(\nu, \omega_{\rm th})= \int_{t_0}^t  \rmd t_1  \int^{\omega_{\rm th}}_0 \rmd \omega \frac{\rmd I}{\rmd \omega \rmd t_1}\,\left[ 1-Q^{\rm NP}_\nu(\omega)\right]\,.
\eeq
Now, assuming full thermalization of such soft modes their distribution should be thermal
\beq
Q^{\rm NP}_\nu(\omega) = \rme^{-\nu (\omega- E^{\rm th}(R,\omega)) }\,.
\eeq
This corresponds to the energy loss function
\beq
S_{\rm loss}^{\rm th}= \delta(\omega-E^{\rm th}(R,\omega))\,.
\eeq
This average energy loss approximation is based on the fact that large number of soft gluons are produced below $\omega_{\rm th}\sim T$ which justifies neglecting fluctuations. The latter are accounted for in the semi-hard sector above $\omega_{\rm th}$.
In the case where the soft gluon energy remaining inside the jet cone is completely thermalized, it can be computed \cite{Mehtar-Tani:2022zwf} and we obtain, 
\beq
E^{\rm th}(R,\omega) =\omega  \left[1+\frac{3}{2}(1+\cos(R))\right]\sin^2\frac{R}{2}\, =\,  \omega R^2+\cO(R^4)\,.
\eeq
This result exhibits the expected scaling of isotropic energy with the jet area, $R^2$. To leading power in $R$, provided $R \ll 1$, medium response can be neglected. Moreover, we are equipped to systematically compute corrections associated with medium response as a power series in $R$.

\section{Towards a higher-order framework}\label{sec:higher-orders}

Let us summarize the main ideas underpinning our factorization approach for jet production based on the large collinear phase space:
\beq
1 \gg R \gg \theta_c\,.
\eeq
The first inequality reflects the standard DGLAP evolution of the jet function while the second corresponds to the non-linear evolution of energy loss due to subjets that are resolved by the medium. At angles of order the coherence angle $\theta_c$ or smaller, interference effects are at work and are power suppressed. A factorization formula that encodes such interferences has ben derived recently \cite{Mehtar-Tani:2024smp}.

The first stage of our factorization involves a scale $\mu$,  such that, $ p_T R  \ll \mu \ll p_T$:
\beq\label{eq:factorization-2}
\frac{\rmd \sigma_{\rm incl}}{\rmd p_T}  = \int_0^1\frac{\rmd x}{x}  \,H\left(\frac{p_T}{x},\mu\right) J\left(x,\frac{p_T}{x},\mu\right)
\eeq
where $H\sim \rmd \sigma_{\rm incl}/\rmd p_T$ is the hard matrix element that can be computed order by order in perturbation theory if one choses $\mu\sim p_T$ and thus does not necessitate any resummation.

On the other hand the jet function involves collinearly enhanced splittings that yield large $\ln R$ that need to be resummed. However, as discussed the Section~\ref{sec:vacuum}, the jet function refactorizes into a linear part that resums DGLAP $\ln R$ and a non-linear part that is sensitive to fluctuations of energy loss due to the substructure dynamics. The latter, resums powers of $\ln (R/\theta_c)$. 

Going beyond the LL accuracy discussed in this paper is straightforward: in addition to higher order corrections to the evolution equation itself, it requires the introduction of perturbative matching coefficients around each of the physical boundaries: $R$ and $\theta_c$ while dropping power suppressed terms of $\cO(\theta_c/R)$ (see also \cite{Mehtar-Tani:2024smp}).

The matching at $\theta \sim R$ is governed by hard vacuum physics and is captured through an expansion in the number of subjets resolved at the hard-collinear scale, $\mu_{\rm hc} \ll p_T R$. The general expression for the factorized jet function, formulated as a product of a hard matching coefficient and energy loss functions summed over the resolved subjets, is given by \cite{Mehtar-Tani:2024smp}:
\beq\label{eq:jet-funct}
&&J\left(x,\frac{p_T}{x},\mu\right) = \,\int_0^{+\infty} \rmd \epsilon\,  \prod_{i=1}^m \int_{\epsilon_i,z_i} C_m\left(\{n_i,z_i\};x'=\frac{p_T+\epsilon}{p_T}x, \frac{\mu}{\mu_{\rm hc}},R\right) \nn
&&\times \,  S_{{\rm loss},m}\left( \epsilon_m, z_m(p_T+\epsilon),\mu_{\rm hc},R\right) \, \delta(\epsilon-\epsilon_1-...-\epsilon_m)\delta(1-z_1-...-z_m)+\cO(R^2)\,,\nn
&& = \,\int_0^{+\infty} \rmd \epsilon\, C_1\left(x'=\frac{p_T+\epsilon}{p_T}x, \frac{\mu}{p_T R}\right) \,  S_{{\rm loss},1}\left( \epsilon, p_T+\epsilon,\mu_{\rm hc}=p_T R,R\right) \,+\cO(\alpha_s) +\cO(R^2)\,.\nn
\eeq
The factorization scales can be chosen to be, $\mu \sim p_T$ and $\mu_{\rm hc} \sim p_T R$. The subscript $m$ represents the number of resolved subjets and $\{n_i,z_i\} $ denotes the energy fraction $z_i$ and the direction $n_i$ of the subjet $i$. While the matching coefficients $C_m(x')$ evolve according to DGLAP evolution with respect to $\mu$, they simultaneously undergo a non-linear DGLAP evolution with respect to $\mu_{\rm hc}$, matching  that of $S_{\rm loss}(\epsilon)$, as dictated by RG consistency. The initial condition for the RG evolution of  the energy loss distribution $S_{\rm loss}(\epsilon)$ is determined by a non-linear rate equation, \eqn{eq:non-lin-eq}, that incorporates the effects of jet energy loss in the medium. The last line in \eqn{eq:jet-funct} depicts the leading log result discussed in this paper so long as $\mu_{\rm hc}= p_T R$.

Of course, if one chooses $\mu=\mu_{\rm hc}$, we would have $C_1(x')=\delta(1-x')$ at leading order and thus,
\beq
J\left(x,\frac{p_T}{x},\mu,R\right) = \,\int_0^{+\infty} \rmd \epsilon\,\delta\left(1-\frac{p_T+\epsilon}{p_T}x \right)  \,  S_{\rm loss}\left( \epsilon, p_T+\epsilon,\mu\right) \,+\cO(R^2)\,,\nn
\eeq
On the other hand, in vacuum we have $S_{\rm loss}(\epsilon)=\delta(\epsilon)$ and as a result we must have 
\beq
J_{\rm vac}\left(x,\frac{\mu}{p_T R}\right) = C_1\left(x'=x, \frac{\mu}{p_T R}\right)\,,\nn
\eeq
In Laplace space \eqn{eq:jet-funct} yields
\begin{align}
J\left(x,\frac{p_T}{x},\frac{\mu}{p_T R},R\right) = p_T \,\int_0^{1} \frac{\rmd z}{z^2} \, {C_1}\left(\frac{x}{z}, \frac{\mu}{\mu_{\rm hc}}\right)  \,  \int \frac{\rmd \nu}{2\pi i } \, \rme^{\left(\frac{1-z}{z}\right) p_T \nu} Q_\nu\left(\frac{p_T}{z},\mu_{\rm hc }\right)+\cO(R^2) \,\,.\nn
\end{align}
Note that, the quenching factor  $Q_\nu$ is implicitly a function of the jet cone size $R$.

At leading log accuracy, by choosing $\mu=p_T$ and $\mu_{\rm hc}\sim p_T R$, we  ensures that the $\ln 1/R$ are resummed in the function $C(x/z,\mu/(p_TR))$ and the $\ln R/ \theta_c$ in the non-linear DGLAP equation
\begin{align}\label{eq:non-lin-eq-coll-3}
& \mu \frac{\del }{\del \mu}Q_\nu(p_T,\mu)  =  \frac{\alpha_s}{\pi} \int_0^1 \rmd z \,p_{gg}(z) \, \left[ Q_\nu(zp_T,\mu)    Q_\nu((1-z)p_T,\mu)  -  Q_\nu(p_T,\mu)   \right]  \,.
\end{align}
At this point, we may simply use this universal evolution with an initial condition fitted to data at an initial scale $\mu_0$ such that all of the medium physics, in particular the phase space constraint are encoded in $Q_\nu(p_T,\mu_0)$. 

At the opposite end of the DGLAP evolution, when $\mu$ approaches the medium scales, order-by-order matching must be performed by computing the collinear medium-induced splitting with full kinematics, including finite-length contributions, which account for the boundaries $\mu \sim z(1-z)\theta_c$ and $\mu \sim \big(z(1-z)\hat{q}\big)^{1/4}$ (see Fig.~\ref{fig:Lund-Plane}).

\section{Conclusion}\label{sec:concl}
In this work, we developed a comprehensive analytic framework for computing the inclusive jet spectrum in heavy-ion collisions, with a focus on the non-linear dynamics driven by medium-induced parton cascades. Our approach captures the intricate interplay between collinear vacuum cascades and turbulent gluon cascades initiated by medium interactions, offering valuable insights into the redistribution of energy both within and outside the jet cone. By systematically resumming leading contributions in parameters such as $\alpha_s \ln(1/R)$, $\alpha_s \ln(R/\theta_c)$, and powers of $\alpha_s L$, our framework extends the factorization derived in \cite{Mehtar-Tani:2024smp} and generalizes the non-linear evolution equations for jet quenching proposed in \cite{Mehtar-Tani:2017web}. Furthermore, we highlighted the dual role of the medium in jet evolution: resolving substructure fluctuations while inducing suppression of micro-jets within the jet cone. This framework, accounts for coherence effects, medium response, and Sudakov suppression in jet quenching.

Our formulation provides a basis for improving precision studies of jet quenching, allowing for the inclusion of higher-order corrections and non-perturbative effects. In addition to inclusive jet cross-sections, it may be applicable to jet substructure observables, enabling a more differential probe of in-medium jet modifications. This new framework offers a promising path forward for future numerical simulations and phenomenological applications, advancing our understanding of jet quenching in heavy-ion collisions.

\section*{Acknowledgements}

We sincerely thank J.-P. Blaizot, P. Caucal, E. Iancu, F. Ringer, B. Singh, A. Takacs and V. Vaidya for their valuable discussions and insightful comments. We extend special thanks to K. Tywoniuk for the many discussions and collaboration on related topics that greatly influenced this work, and to P. Caucal for his meticulous review of the manuscript. This work was supported by the U.S. Department of Energy under Contract No. DE-SC0012704 and by Laboratory Directed Research and Development (LDRD) funds from Brookhaven Science Associates.


\bibliographystyle{JHEP}
\bibliography{paper-jet-eloss}

\end{document}